\documentclass[preprint,12pt]{elsarticle}

\makeatletter
\def\ps@pprintTitle{%
 \let\@oddhead\@empty
 \let\@evenhead\@empty
 \def\@oddfoot{}%
 \let\@evenfoot\@oddfoot}
\makeatother

\usepackage{amssymb}

\usepackage{float}
\usepackage{subfigure}
\usepackage{amsmath}
\usepackage{afterpage}
\usepackage{longtable}

\usepackage{graphicx,subfigmat,etoolbox,amssymb,float}

\newcounter{subfigcount}
\begin{document}

\begin{frontmatter}



\title{Aeroacoustic Source Localization}

 \author[label1,label2]{Weicheng Xue}
 \affiliation[label1]{organization={Department of Computer Science and Technology, Tsinghua University},
             city={Beijing},
             postcode={100084},
             country={China}}

 \author[label2]{Bing Yang}
 \affiliation[label2]{organization={Institute of Engineering Thermophysics, Chinese Academy of Sciences},
             city={Beijing},
             postcode={100083},
             country={China}}             

 \author[label2]{Shaohong Jia}


\begin{abstract}
The deconvolutional DAMAS algorithm can effectively eliminate the misconceptions in the usually-used beamforming localization algorithm, allowing for more accurate calculation of the source location as well as the intensity. When solving a linear system of equations, the DAMAS algorithm takes into account the mutual influence of different locations, reducing or even eliminating sidelobes and producing more accurate results.

This work first introduces the principles of the DAMAS algorithm. Then it applies both the beamforming algorithm and the DAMAS algorithm to simulate the localization of a single-frequency source from a 1.5 MW wind turbine, a complex line source with the text "UCAS" and a line source downstream of an airfoil trailing edge. Finally, the work presents experimental localization results of the source of a 1.5 MW wind turbine using both the beamforming algorithm and the DAMAS algorithm.
\end{abstract}


\begin{highlights}
\item This work applied the beamforming algorithm and the DAMAS algorithm to perform sound source localization for numerical simulations as well as real-world experiments for a 1.5 MW wind turbine.
\item The results presented in this work showed the comparison of the beamforming algorithm and the DAMAS algorithm and displayed the effectiveness and accuracy of the DAMAS algorithm. The DAMAS algorithm can calculate the position and intensity of the sound source more accurately, with smaller main lobe sizes in the resulting intensity maps.
\item The diagonal-removed beamforming algorithm and the DAMAS algorithm can be combined for a more accurate understanding of sound source distribution and for further noise reduction methods in the domain of wind turbine noise reduction.
\end{highlights}

\begin{keyword}
DAMAS \sep beamforming \sep source localization \sep experiment \sep simulation


\end{keyword}

\end{frontmatter}


\section{Introduction}

The noise radiation from wind turbines can potentially or has already been proven to have significant impacts on the ecological environment and residents' living conditions. Knowing this, an investigation of the mechanisms of noise generation and the characteristics of wind turbine noise is necessary and urgent. Therefore, the localization of noise sources has always been a hot topic in the field of fluid acoustics. For large-scale wind turbines, aerodynamic noise and nacelle noise are potential primary noise sources. Therefore, accurate localization of noise sources can promote the further development of effective measurement for noise reduction for wind turbines and promote research on the mechanisms of aerodynamic noise.

Due to the importance of noise control in industries such as aviation and wind power generation, many researchers have conducted valuable studies on aerodynamic sound localization. Faser et al.~\cite{faszer2006acoustic} conducted a series of sound source localization experiments on a number of airfoil models with different trailing edges exploiting both the beamforming and SONAH algorithms. They found that the beamforming algorithm can result in a larger main lobe size and lower intensity of the noise source, compared with the SONAN algorithm. However, the SONAN algorithm can predict the noise source location more accurately as well as the intensity of the noise source at low frequencies. For high-frequency noise sources, the SONAN algorithm did not perform well due to the limitation of the number of microphones used. Sijtsma et al.~\cite{sijtsma2001location} derived formulas for the localization of moving sound sources and experimentally validated the formulas using a rotating helicopter wing model and a wind turbine blade model. The beamforming algorithm can produce a clear visualization of the rotating noise source in their helicopter wing experiment and a good identification of the leading-edge and trailing-edge noise sources in their wind turbine blade experiment. Oerlemans et al.~\cite{oerlemans2005acoustic} also conducted aerodynamic sound source localization experiments on a full-scale wind turbine using the beamforming algorithm. They obtained and analyzed the localization results under different wind speeds, power levels, yaw angles, rotational speeds, and pitch angles. Besides, Oerlemans et al.~\cite{oerlemans2007quantification} conducted noise source localization on an A340 model in both closed and open wind tunnels. They found that the noise source characteristics were similar when the flow conditions were similar. However, due to the limitation of the number and arrangement of the microphones in an array, results generated from the traditional beamforming algorithm may lead to ambiguous or even erroneous interpretations. To eliminate the interference of sidelobes and obtain higher-resolution sound intensity maps, Pieter Sijtsma et al.~\cite{sijtsma2007clean} proposed the CLEAN-SC algorithm. Using the CLEAN-SC algorithm, more accurate sound source intensity information and distribution can be obtained, as demonstrated in experiments with the A340 model in a closed wind tunnel. Brooks et al. [6] proposed the DAMAS algorithm, a deconvolution algorithm that can generate accurate localization results. They conducted a series of numerical verifications and experimental validations using their proposed DAMAS algorithm. Compared with the traditional beamforming algorithm, the DAMAS algorithm displayed obvious advantages in the sidelobe suppression and main lobe enhancement. Moreover, the localization results obtained using the DAMAS algorithm can be more intuitive to understand.

The work first introduces the basic principles of the DAMAS algorithm~\cite{brooks2006deconvolution}. Then, it applies both the beamforming algorithm and the DAMAS algorithm to simulate the localization of a single-frequency source from a 1.5 MW wind turbine, a complex line source with the text "UCAS" and a line source downstream of an airfoil trailing edge. By comparing the results of the two algorithms, it is clear that the DAMAS algorithm has an advantage over the beamforming algorithm in terms of accuracy in calculating the noise source location as well as the intensity. Finally, the work presents experimental localization results on the noise source of a 1.5 MW wind turbine using both the beamforming algorithm and the DAMAS algorithm. The DAMAS algorithm can assist to reduce misconceptions and misjudgments regarding the distribution of the wind turbine's noise sources. In summary, this work concludes that sound source localization can benefit from the combined use of the beamforming algorithm with the diagonal removal and the DAMAS algorithm, which enables a more accurate understanding of the noise source distribution and can facilitate future technology of noise reduction.

\section{The Principle of DAMAS}

The first step involves scanning the grid surface of suspicious noise sources using the beamforming algorithm. Assuming the signals from each microphone are represented as $p_m(t) (m = 1, 2, ..., m_0)$, where $t$ is the time domain and $m_0$ is the total number of microphones used, we transform these signals from the time domain to the frequency domain to obtain $p_m(f)$. The microphone vector is denoted in Eq.~\ref{array_vector}:

\begin{equation}
P = col[p_1(f)\ p_2(f)\ ...\ p_{m_0}(f)]
\label{array_vector}
\end{equation}

Based on the frequency domain signals of all the microphones, the cross-spectrum matrix can be obtained in Eq.~\ref{cross_spectrum}:

\begin{equation}
PP = P^T P
\label{cross_spectrum}
\end{equation}
where the superscript $T$ denotes a complex transpose. It is straightforward to know that the shape of the cross-spectrum matrix $PP$ is $m_0 \times m_0$, and the lower triangle elements are the complex transpose of the upper triangle elements in the cross-spectrum matrix $PP$. 

The control function connecting each grid point on the noise source surface to each microphone on the array surface is given in Eq.~\ref{control_func}, and the vector according to the numbering of the microphones for each control function is represented in Eq.~\ref{control_func_vec}:

\begin{equation}
e_m = \frac{r_m}{r_c} e^{j 2 \pi f \Delta t_m} \ (m = 1, 2, ..., m_0)
\label{control_func}
\end{equation}

\begin{equation}
E = col[e_1\ e_2\ ...\ e_{m_0}]
\label{control_func_vec}
\end{equation}
where $r_c$ is the distance from the center of the microphone array to the noise source surface. $\frac{r_m}{r_c}$ is used to normalize the distance between each grid location and each microphone. The phase in the control function is defined in Eq.~\ref{phase}:

\begin{equation}
2 \pi f \Delta t_m = \vec{k}\Dot{\vec{x_m}} + 2 \pi f \Delta t_{m,shear}
\label{phase}
\end{equation}
where $\Delta t_{m,shear}$ is the temporal latency caused by the flow shear layer.

The power spectrum matrix of the traditional beamforming algorithm is given in Eq.~\ref{power_spectrum}:

\begin{equation}
Y = \frac{E^T PP E}{m_0^2}
\label{power_spectrum}
\end{equation}

In addition, the elements of the cross spectrum matrix in Eq.~\ref{cross_spectrum}, representing the individual microphones, can be subjected to a weighted operation. The new cross spectrum matrix can be obtained as $PP_{new} = W PP W^T$, where $W$ is a weight matrix given in Eq.~\ref{weight}.

\begin{equation}
W = 
\begin{bmatrix}
w_1 & 0 & & ... & 0 \\
0 & w_2 & & ... & 0 \\
  &     & ... \\
0 & 0 & & ... & w_{m_0}
\end{bmatrix}
\label{weight}
\end{equation}
where $w_m$ is the weight for the $m-th$ microphone. When measuring low frequency signals, weights for the inner-ring microphones in an array can be set to 0. When measuring high frequency signals, weights for the outer-ring microphones can be set to 0.

A method to remove the array's autocorrelation spectral noise is by setting the elements on the diagonal of $PP$ to 0. However, this may result in negative sound intensity in the low-noise grid region, which obviously contradicts the real physical scenario. Considering the shortcoming, the DAMAS algorithm used in this study does not employ the diagonal removal method.

It should be noted that when a weighted or diagonal-removed cross-power spectrum matrix is used, the corresponding microphone weights should also be subtracted from the denominator.

\section{DAMAS Inverse Problem}

The cross-spectrum matrix caused by grid $n$ in the DAMAS algorithm is given in Eq.~\ref{cross_matrix_n}:

\begin{equation}
G_n = X_n
\begin{bmatrix}
(e_1^(-1))^* e_1^(-1) & (e_1^(-1))^* e_2^(-1) & ... & (e_1^(-1))^* e_{m_0}^(-1) \\
(e_2^(-1))^* e_1^(-1) & (e_2^(-1))^* e_2^(-1) & ... & (e_2^(-1))^* e_{m_0}^(-1) \\
 & & ... & \\
(e_{m_0}^(-1))^* e_1^(-1) & (e_{m_0}^(-1))^* e_2^(-1) & ... & (e_{m_0}^(-1))^* e_{m_0}^(-1) 
\end{bmatrix}
\label{cross_matrix_n}
\end{equation}

where $X_n$ is the square of the sound pressure caused by the grid $n$ supposing the microphone is placed at the center of the array. If there are $N$ grid points involved, the total cross-power spectrum matrix is given in Eq.~\ref{total_cross_spectrum}:

\begin{equation}
G = \sum_{n=1}^{N} {G_n}
\label{total_cross_spectrum}
\end{equation}

Therefore, the sound intensity at each point on the grid plane obtained through the DAMAS algorithm is given in Eq.~\ref{Y_n}:

\begin{equation}
Y_n = [\frac{e^T G e}{m_0^2}]_n = \sum_{n^{'}=1}^{N} {\frac{e^T G_{n^{'}} e}{m_0^2} X_{n^{'}}}
\label{Y_n}
\end{equation}
Let
\begin{equation}
A_{n n^{'}} = {\frac{e^T G_{n^{'}} e}{m_0^2}}
\label{Ann}
\end{equation}
By setting $Y_n$ equal to the measured $Y_n$, a system of equations can be obtained and is given in Eq.~\ref{system_equation}:

\begin{equation}
A X = Y
\label{system_equation}
\end{equation}
where $A$, $X$, and $Y$ have elements $A_{nn^{'}}$, $X_n$, and $Y_n$, respectively. Similarly, weighted approaches can also be used in the DAMAS algorithm, including diagonal the removal and other techniques.

Let us take a look at the characteristics of the coefficient matrix $A$ in Eq.~\ref{system_equation}. When calculating $G$ or $G_n$, if no weighting or diagonal removal is applied, all the diagonal elements of $A$ are unity, and all the elements are positive. If diagonal removal is applied, the main diagonal elements of $A$ are null. However, the off-diagonal elements may be negative if the grid points represented by $n$ and $n^{'}$ are sufficiently far apart. Understanding these can be highly useful for solving the system of equations later in this work.

\section{DAMAS Algorithm}

Only in a few rare scenarios, the matrix $A$ in the DAMAS algorithm is not a singular matrix. In specific domains, such as in aerodynamic acoustic sound source localization, the rank of matrix $A$ may be much lower compared to its number of rows, reaching only 0.25 or even lower. This feature implies that there are an infinite number of solutions to the system of equations. Nevertheless, it is still possible to solve this system of equations using a simple linear iterative formula, and the solution can converge rapidly. The DAMAS algorithm takes advantage of this property and iteratively updates the sound source distribution until the convergence tolerance is hit. The relatively rapid convergence of the DAMAS algorithm makes it an effective method for solving the system of equations and obtaining reasonable solutions in the domain of aerodynamic acoustic sound source localization.

To display the iterative process, Eq.~\ref{system_equation} can be expanded and rewritten in Eq.~\ref{system_equation_expanded}.

\begin{equation}
A_{n1} X_1 + A_{n2} X_2 + ... + A_{nn} X_n + A_{nN} X_N = Y_n
\label{system_equation_expanded}
\end{equation}
Let $Ann^{’} = 1$, the resulting iterative equations for DAMAS algorithm can be expressed in Eq.~\ref{system_equation_iterative}:
\begin{equation}
X_n = Y_n - [\sum_{n^{'}=1}^{n-1} {A_{nn^{'}} X_{n^{'}}} + \sum_{n^{'}=n+1}^{N} {A_{nn^{'}} X_{n^{'}}}]
\label{system_equation_iterative}
\end{equation}

The initial value of $X_n$ for the first iteration can be set as either 0 or $Y_n$. If any value of $X_n$ becomes negative during the iteration process, it is set to 0 to ensure non-negativity. To ensure smooth progression of the iteration, the iteration is performed in a specific order where it starts from grid 1 and progresses to grid $N$, and then reverses from grid $N$ to grid 1. By setting an appropriate initial value and ensuring non-negativity in each iteration step, the algorithm aims to converge towards a solution that satisfies the measured sound pressure distribution and produces a smooth and physically meaningful sound source distribution.

\section{Numerical Verification}

\subsection{Single-frequency Noise Source for a Wind Turbine}

A UP77/1500IAALT wind turbine generator with a power of 1.5 MW is chosen as the subject for the first numerical simulation. The parameters for the simulation are given in Table~\ref{parameters_windturbine}:

\begin{table}[H]
	\caption{Parameters for the single-frequency noise source for a wind turbine}
	\centering
	\begin{tabular}{cc}
		\hline
		Wind turbine rotor radius& $R = 38.68$ m \\
		Microphone array radius& $R_m = 1$ m\\
		Distance between the array plane and the rotor plane& $D = 65$ m\\
        Distance between the source and the rotor plane center& R0 = 1 m\\
        Sound pressure of the noise source& $p0 = 2$ Pa\\
        Frequency of the noise source& $f0 = 1000$ Hz\\
        Angle with respect to the horizontal plane& $\theta_0 = 45^{\circ}$\\
		\hline
	\end{tabular}
	\label{parameters_windturbine}
\end{table}

These parameters are used to define the specific configuration and characteristics of the simulation scenario for the wind turbine aeroacoustic noise source.

\begin{figure}[H]
	\centering
	\includegraphics[width=1.0\textwidth]{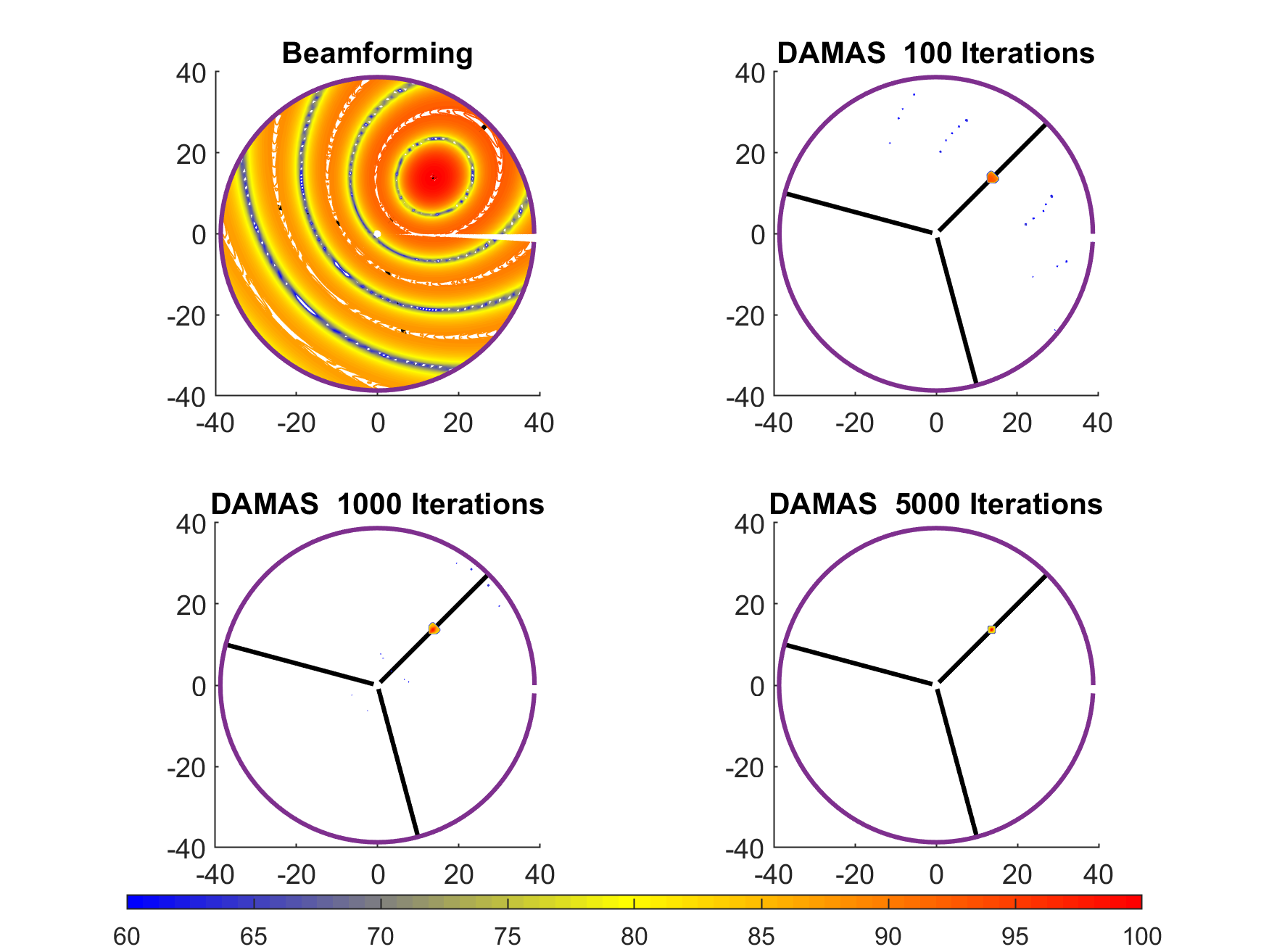}
	\caption{Single-frequency sound source localization for a wind turbine using beamforming and DAMAS}
	\label{former_mill}
\end{figure}

The results of the beamforming algorithm and DAMAS algorithm, along with their respective maximum values and the contour maps within the -40 dB range, are shown in Fig.\ref{former_mill}. The black lines in Fig.\ref{former_mill} denote three blades of the wind turbine. The top-left subfigure represents the results obtained using the beamforming algorithm. It can be observed that even within the -40 dB range, the noise source fills almost the entire area, which is not consistent with the actual conditions. On the other hand, for the DAMAS result after 100 iterations which is the top-right subfigure, the location of the main noise source is accurately identified and the size of the main lobe is small. By summing the noise sources within the main lobe area, the intensity reaches 100.13 dB, which is very close to the actual noise source intensity. However, there are still noises surrounding the main lobe with lower intensities. As the iteration progresses to 1000 steps, the intensity of the main sound source becomes closer to the actual intensity, and the number of false noise sources around the correct location as well as the intensity decreases further. After 5000 iteration steps, the intensity of the main noise source is almost equal to the intensity initially set. Also, there are no other false noise sources within the -40 dB range.

\section{Complex Line Noise Sources on Grid Points}

In this case, the simulation is performed on a complex line noise source with the inscription "UCAS" assuming that the noise sources are located exactly on the grid points. Considering the importance of the ratio $\Delta x / B$ between the grid spacing $x$ and the array resolution size $B$ in precision control, two scenarios are considered: $\Delta x / B = 0.083$ and $\Delta x / B = 0.167$. This work will show that more accurate calculation results can be obtained when choosing a larger value of $\Delta x / B$.

\begin{figure}[H]
	\centering
	\includegraphics[width=1.0\textwidth]{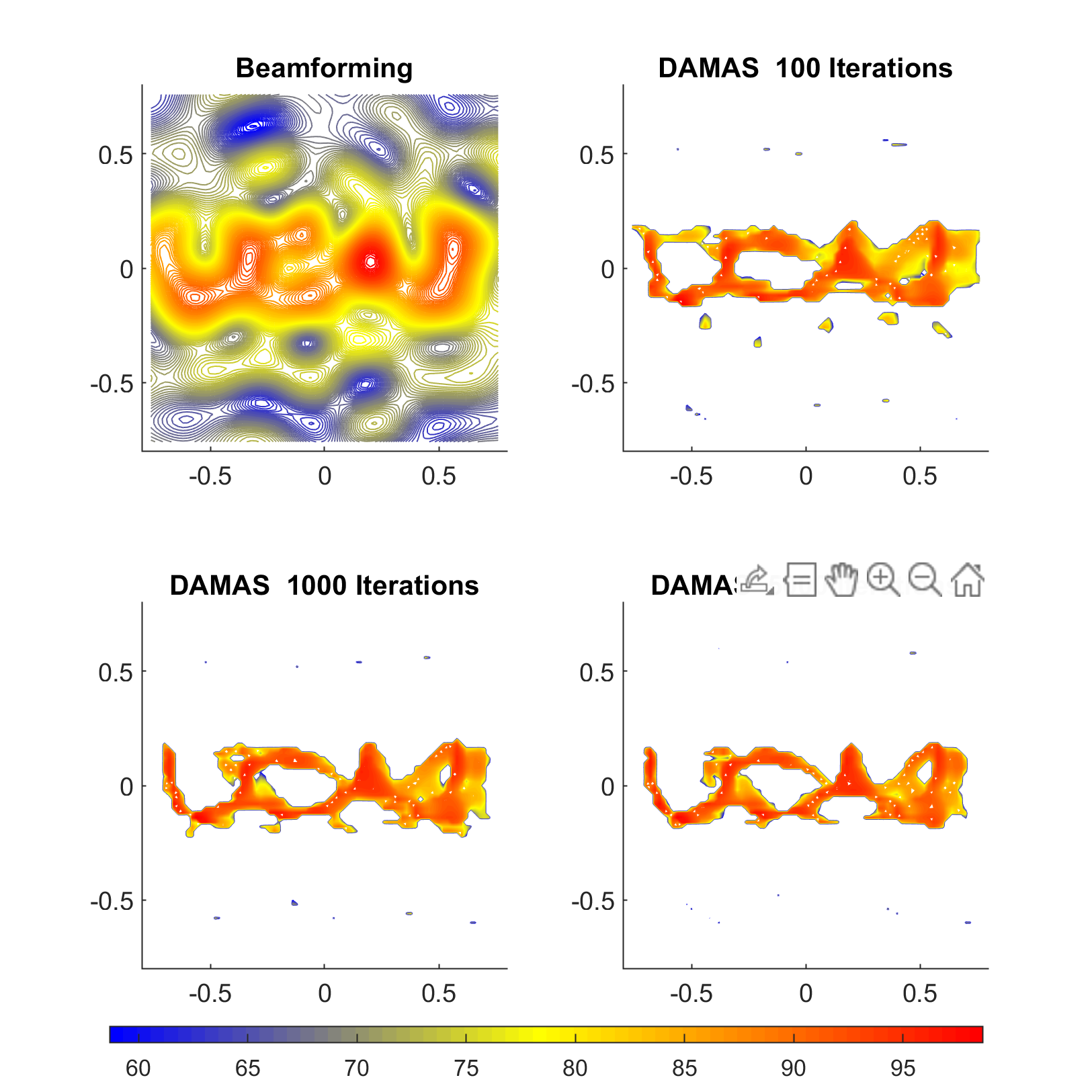}
	\caption{Sound source localization for a "UCAS" inscription on grid points (10000 Hz) using beamforming and DAMAS}
	\label{UCAS_10000}
\end{figure}

\begin{figure}[H]
	\centering
	\includegraphics[width=1.0\textwidth]{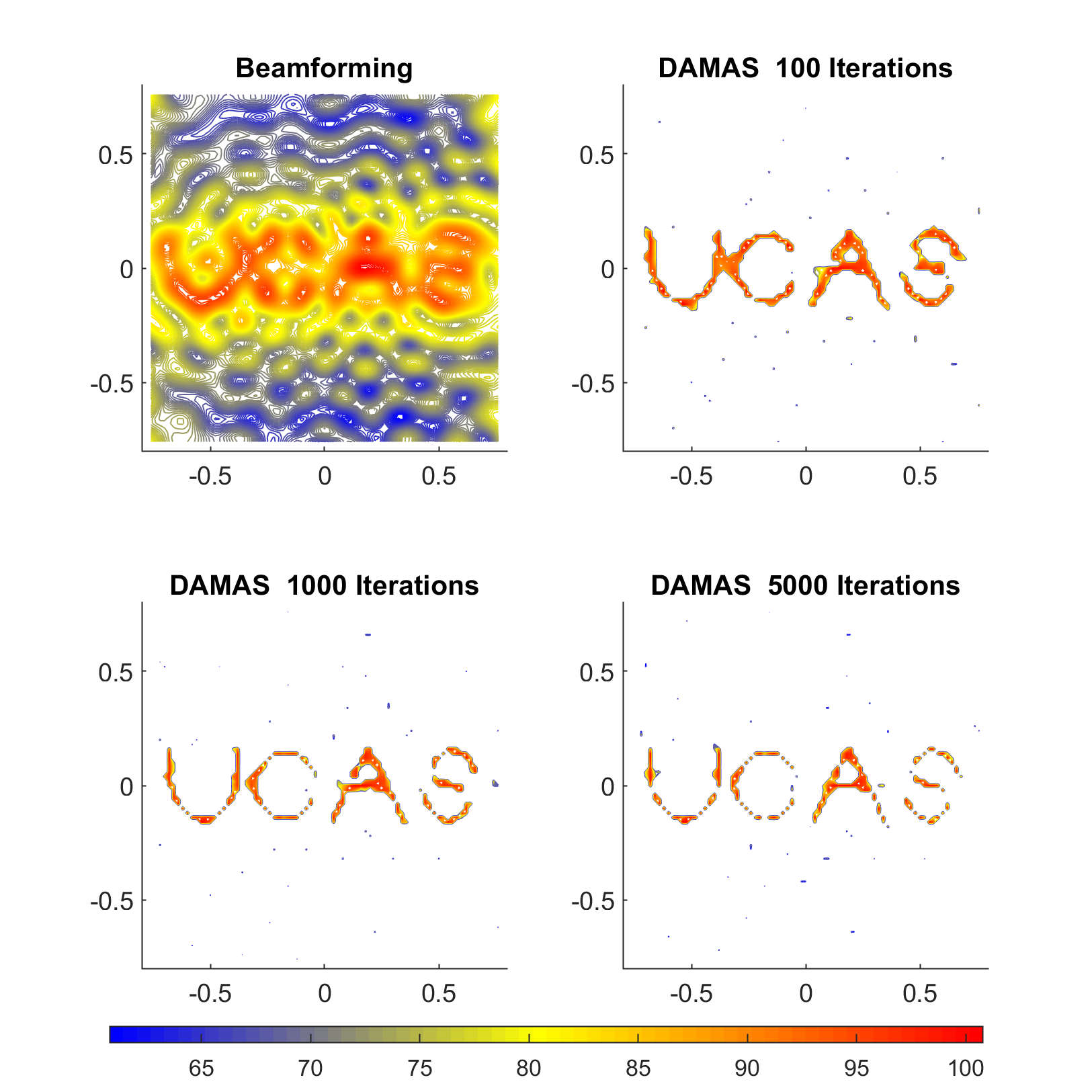}
	\caption{Sound source localization for an "UCAS" inscription on grid points (20000 Hz) using beamforming and DAMAS}
	\label{UCAS_20000}
\end{figure}

Initially, all the intensities at different positions corresponding to the "UCAS" inscription on the grid plane are set to the same, such as 100 dB, while the intensities on other grid points are set to 0. This arrangement precisely sets the profile of the UCAS inscription with line noise sources. From the DAMAS algorithm results shown in Fig.~\ref{UCAS_10000}, the high intensity locations in the contour plot starts to resemble the UCAS inscription ater 100 iterations. After further iterations, the displayed line sources become thicker, making it easier to discern the UCAS inscription. It should be noted that the simulation has not converged even after 5000 iterations, but the accuracy has significantly improved compared to the beamforming algorithm. From the DAMAS algorithm results for the high frequency sound localization shown in Fig.~\ref{UCAS_20000}, the "UCAS" inscription is more clear and recognizable. If the noise source intensities are summed over all grid points, the difference of the calculated result at 100 iterations is only 0.0072 dB different from the initial setting.

\section{complex Line Noise Sources on Grid Points and Non-grid Points}

\begin{figure}[H]
	\centering
	\includegraphics[width=1.0\textwidth]{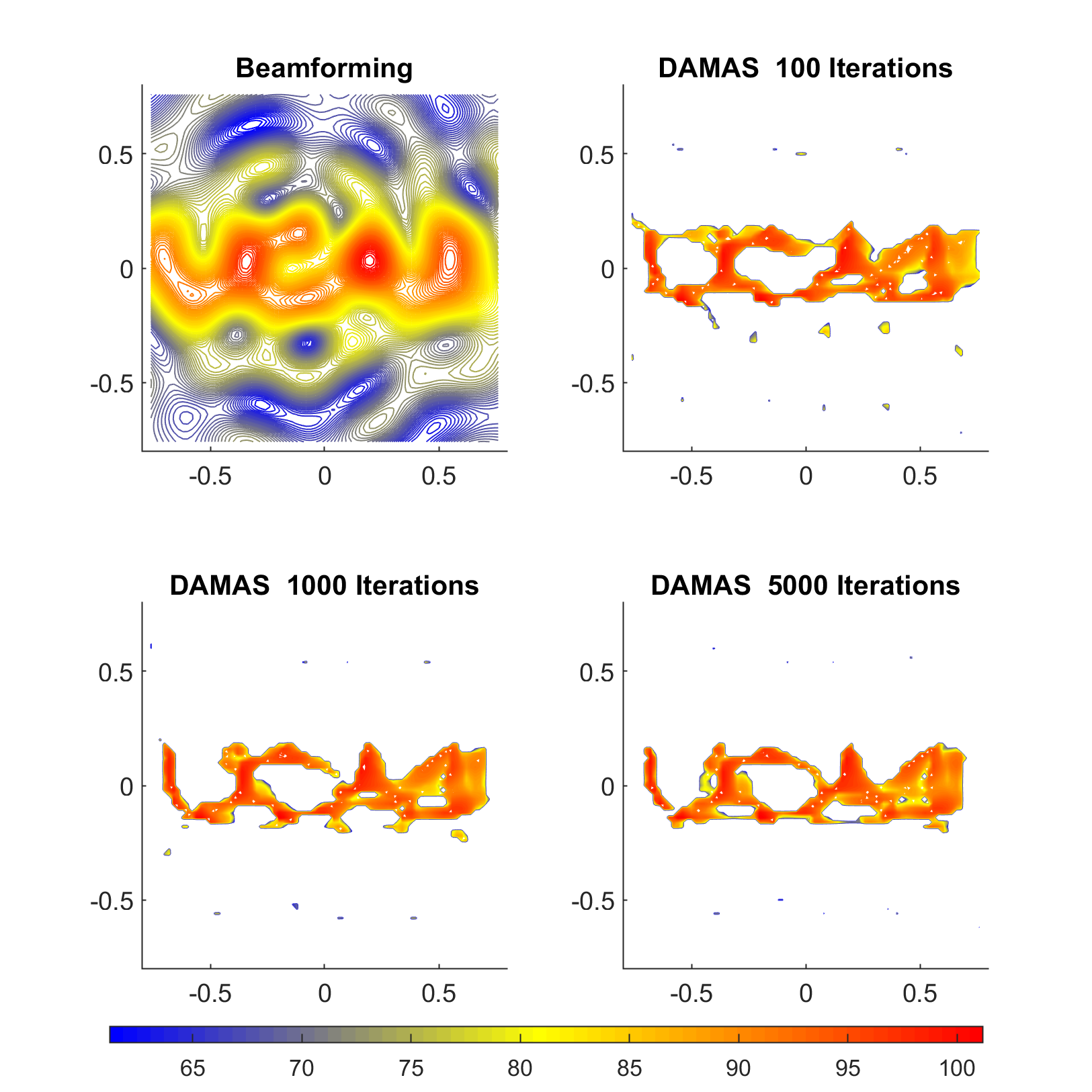}
	\caption{Sound source localization for an "UCAS" inscription on grid points and non-grid points (10000 Hz) using beamforming and DAMAS}
	\label{UCAS_10000_nongrid}
\end{figure}

\begin{figure}[H]
	\centering
	\includegraphics[width=1.0\textwidth]{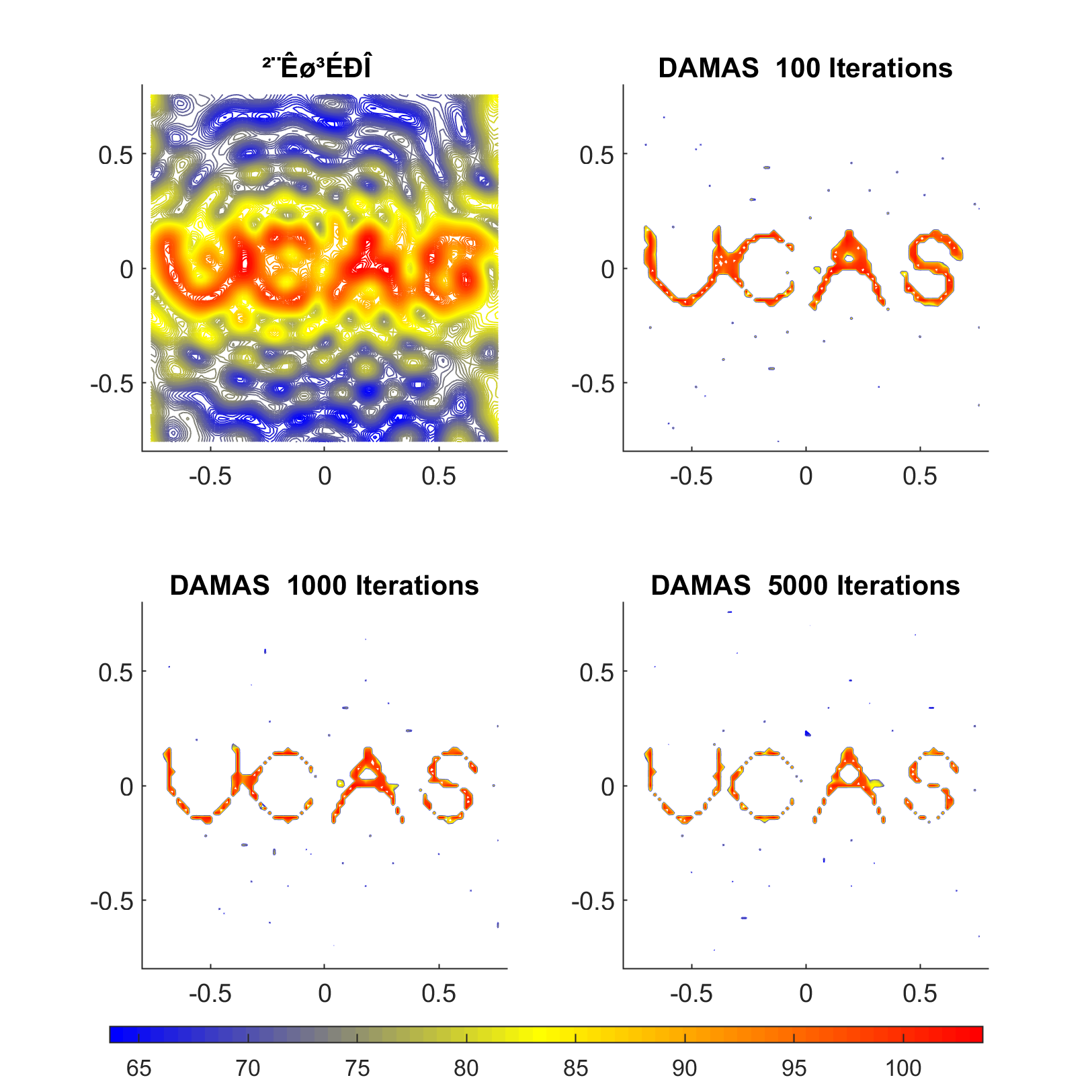}
	\caption{Sound source localization for an "UCAS" inscription on grid points and non-grid points (20000 Hz) using beamforming and DAMAS}
	\label{UCAS_20000_nongrid}
\end{figure}

The comparison of Fig.~\ref{UCAS_10000} and Fig.~\ref{UCAS_20000} indicates that the same level of computational accuracy can be achieved with fewer iteration steps as the value of $\Delta x/B$ increases. However, we are particularly interested in scenarios in which the noise sources not only fall on the grid points but also between the grid points. Fig.\ref{UCAS_10000_nongrid} displays the DAMAS simulation results for $\Delta x/B = 0.083$, where the noise sources are positioned between the grid points. The total noise source intensity differs from the initial intensity by only 0.01 dB after running for 100 iterations, which is slightly larger than the case in which the noise sources are located only at the grid points. Fig.\ref{UCAS_20000_nongrid} shows the DAMAS simulation results for non-grid point complex line noise sources with $\Delta x/B = 0.167$. The total noise source intensity differs from the initial intensity by only 0.016 dB after 100 iterations. The difference in intensity is further reduced to only 0.006 dB after 1000 iterations. These results demonstrate that the DAMAS algorithm can effectively simulate complex line noise sources even when they are positioned between grid points or non-grid points. The choice of $\Delta x/B$ value affects the computational accuracy and convergence speed, and a larger value of $\Delta x/B$ allows for achieving comparable accuracy with fewer iterations.

\section{Simulation of Downstream Aifoil Trailing Edge Line Noise Sources}

\begin{figure}[H]
	\centering
	\includegraphics[width=1.0\textwidth]{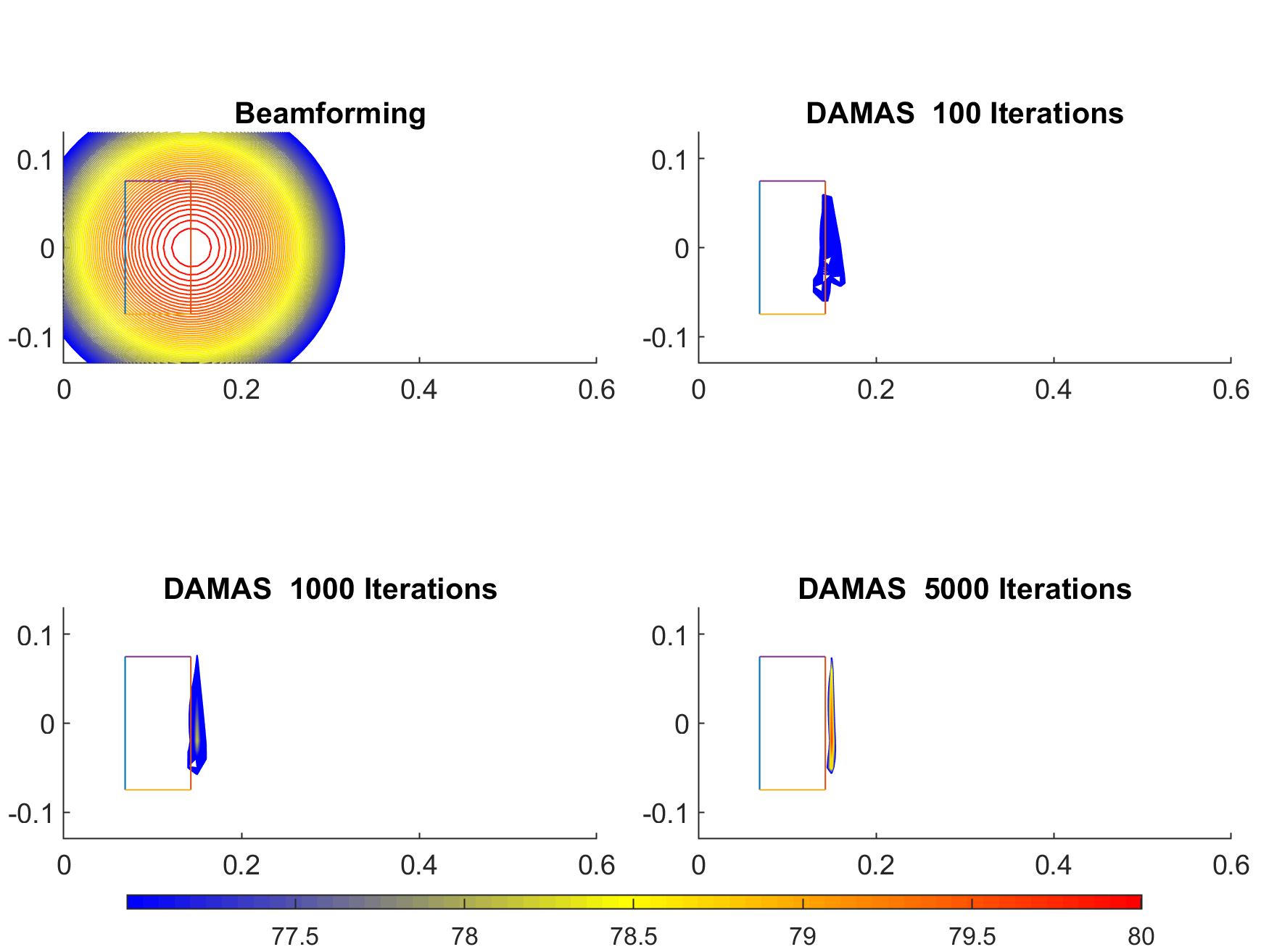}
	\caption{Sound source localization for downstream aifoil trailing edge line noise sources using beamforming and DAMAS}
	\label{airfoil}
\end{figure}

Fig.\ref{airfoil} shows the numerical simulation of the airfoil trailing edge noise experiment conducted in an anechoic wind tunnel. The real problem is simplified to a line noise source with a length equal to the airfoil span located downstream of the airfoil trailing edge. The frequency of the noise source is set to be 2000 Hz and the intensity is set to be 80 dB. From the result obtained using the beamforming algorithm in the top left subfigure, it can be observed that due to the small size of the wind tunnel and the airfoil, the main lobe width (-3 dB) region is relatively large, even encompassing the entire airfoil, which is highly undesirable. After 100 iterations using the DAMAS algorithm, the main lobe size has significantly reduced and can be used for precise localization of airfoil trailing edge noise. As the number of iterations increases, the intensity of the sound source approaches to the correct locations, with a difference of only 0.0082 dB at 1000 iterations, and no interfering noise is displayed. However, the DAMAS algorithm still has limitations as the noise source intensities on various grid points of the line source are not consistent and do not fully match the initial conditions. At 1000 iterations, there is a difference of 30 dB between the strongest and weakest noise points on the trailing edge of the airfoil. At 5000 iterations, the difference between the strongest and weakest noise points is still approximately 24 dB. Nevertheless, compared to the beamforming algorithm, the DAMAS algorithm has significantly improved accuracy.

\section{Experimental Study on Sound Source Localization of a 1.5 MW Wind Turbine}

\begin{figure}[H]
\begin{longtable}{@{}ccc@{}}
    {\includegraphics[width=.3\linewidth,trim=66 6 66 6,clip]{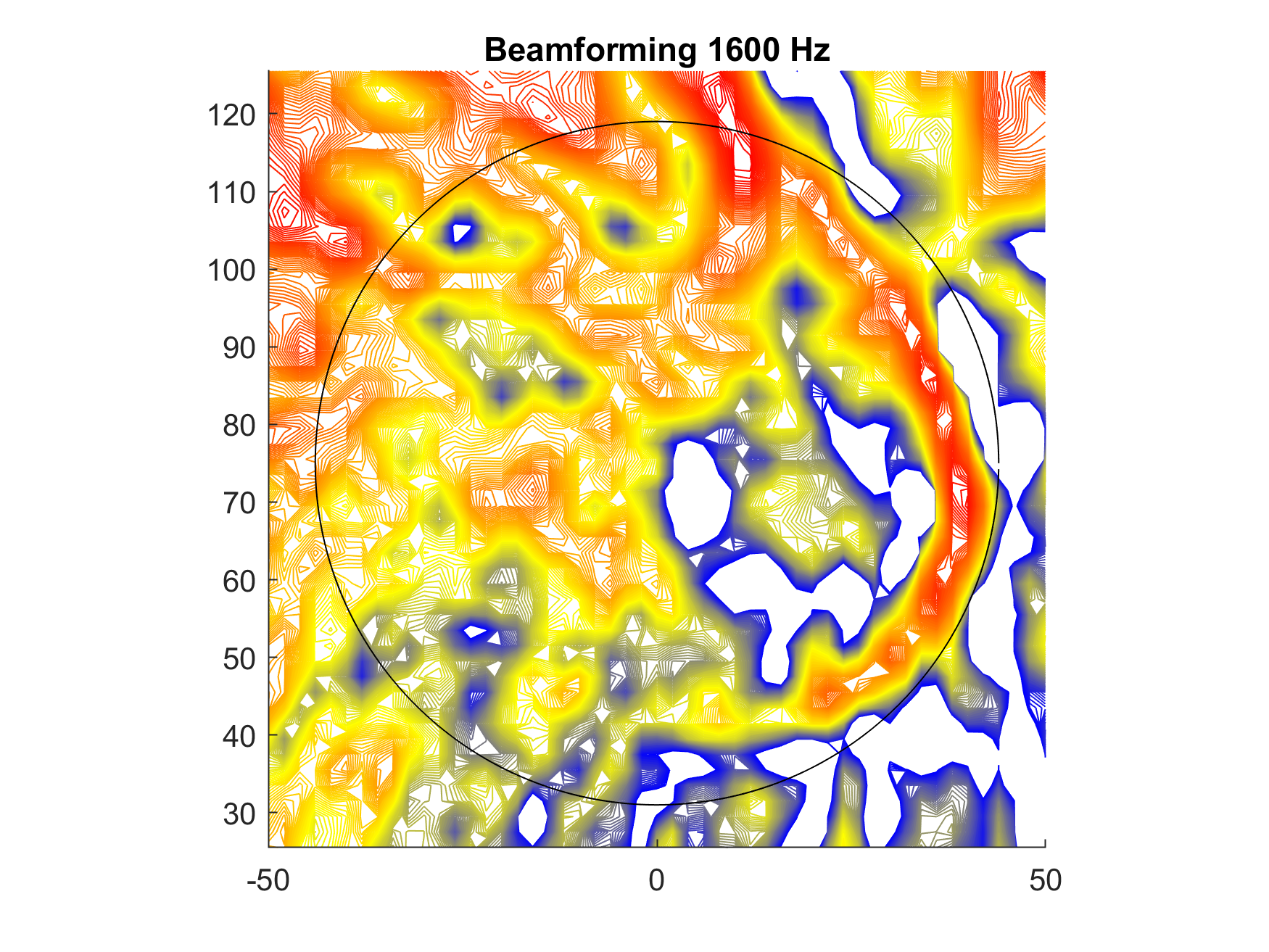}} &   
    {\includegraphics[width=.3\linewidth,trim=66 6 66 6,clip]{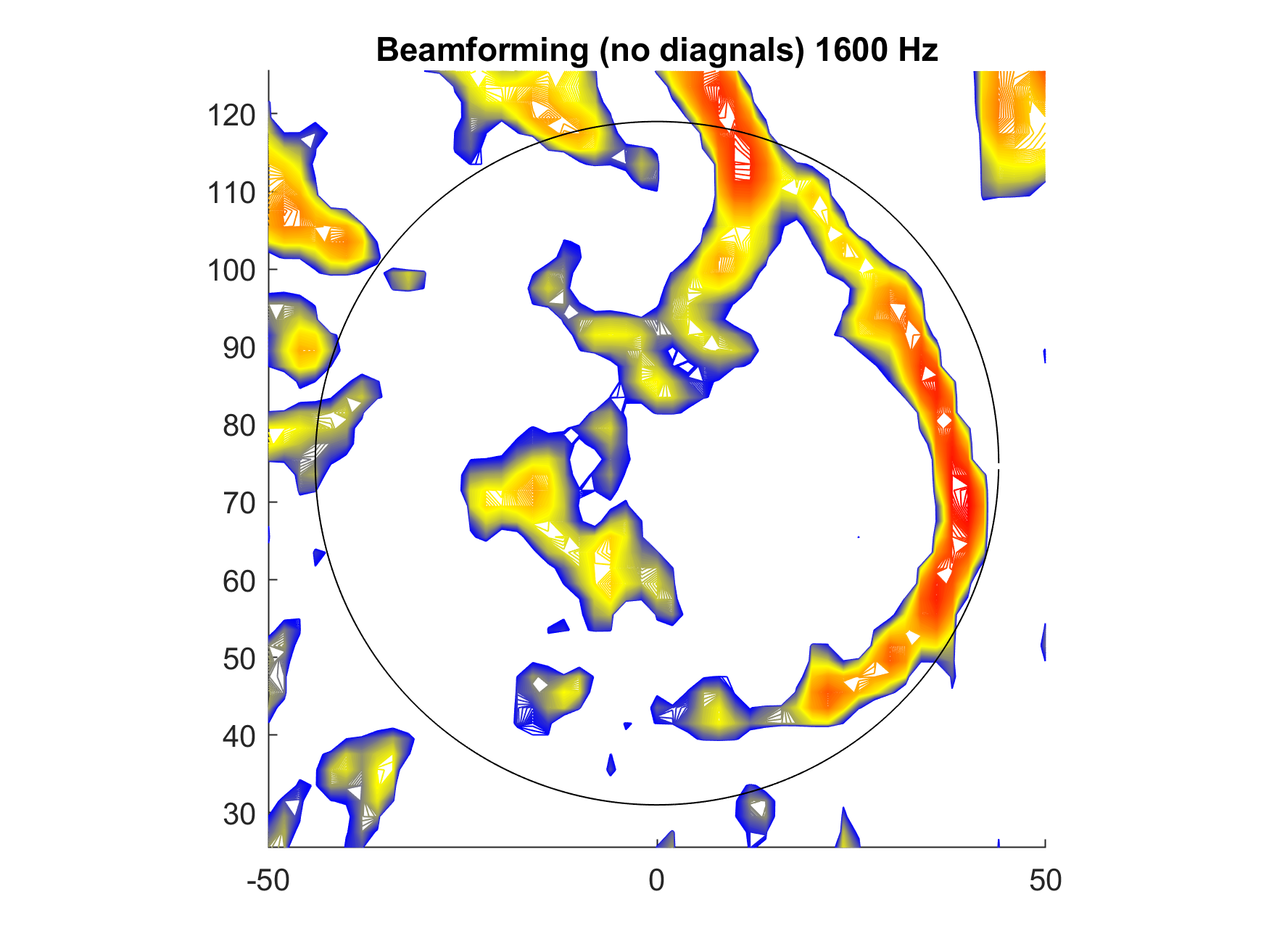}} &   
    {\includegraphics[width=.3\linewidth,trim=66 6 66 6,clip]{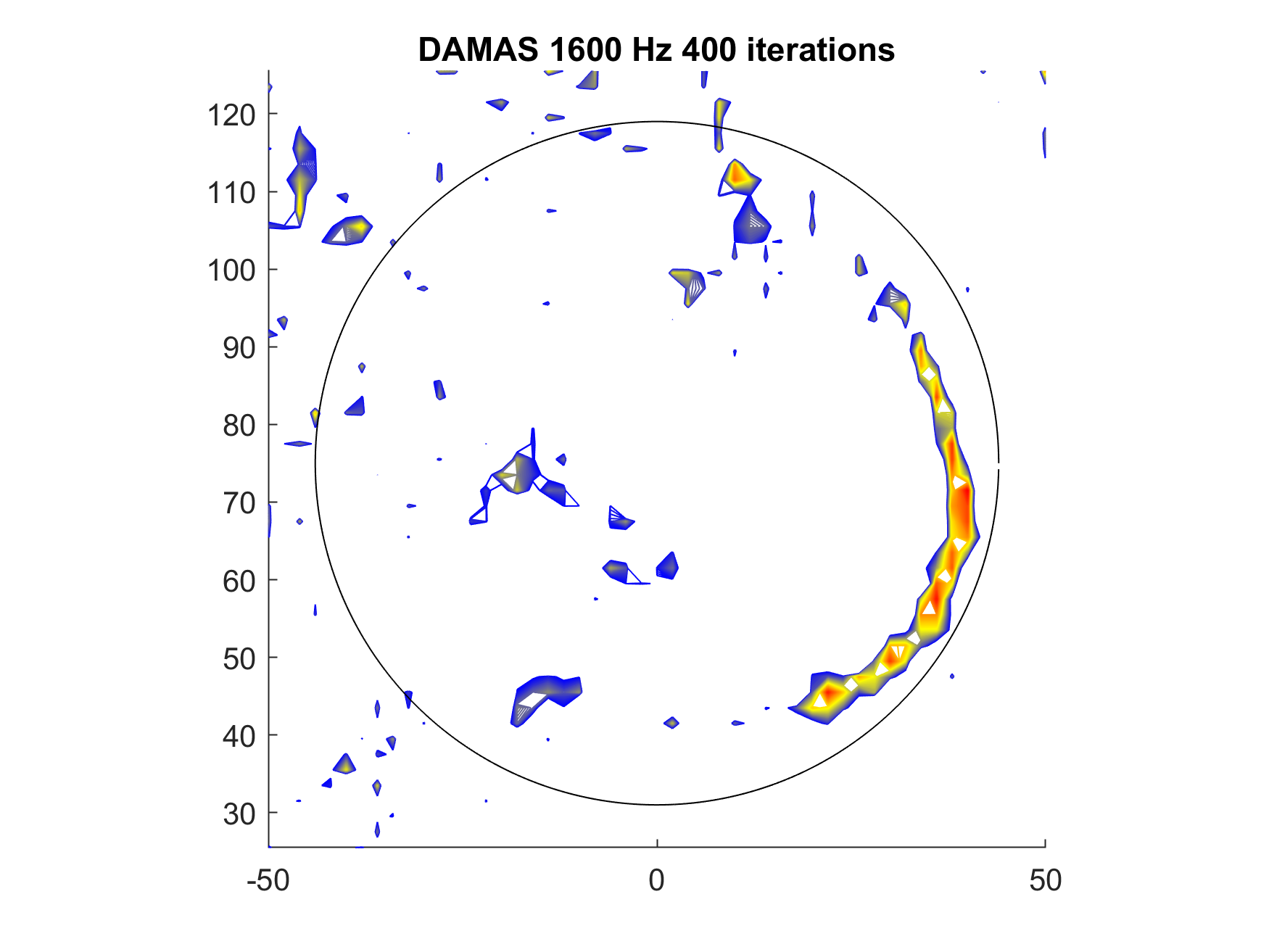}} \\
    {\includegraphics[width=.3\linewidth,trim=66 6 66 6,clip]{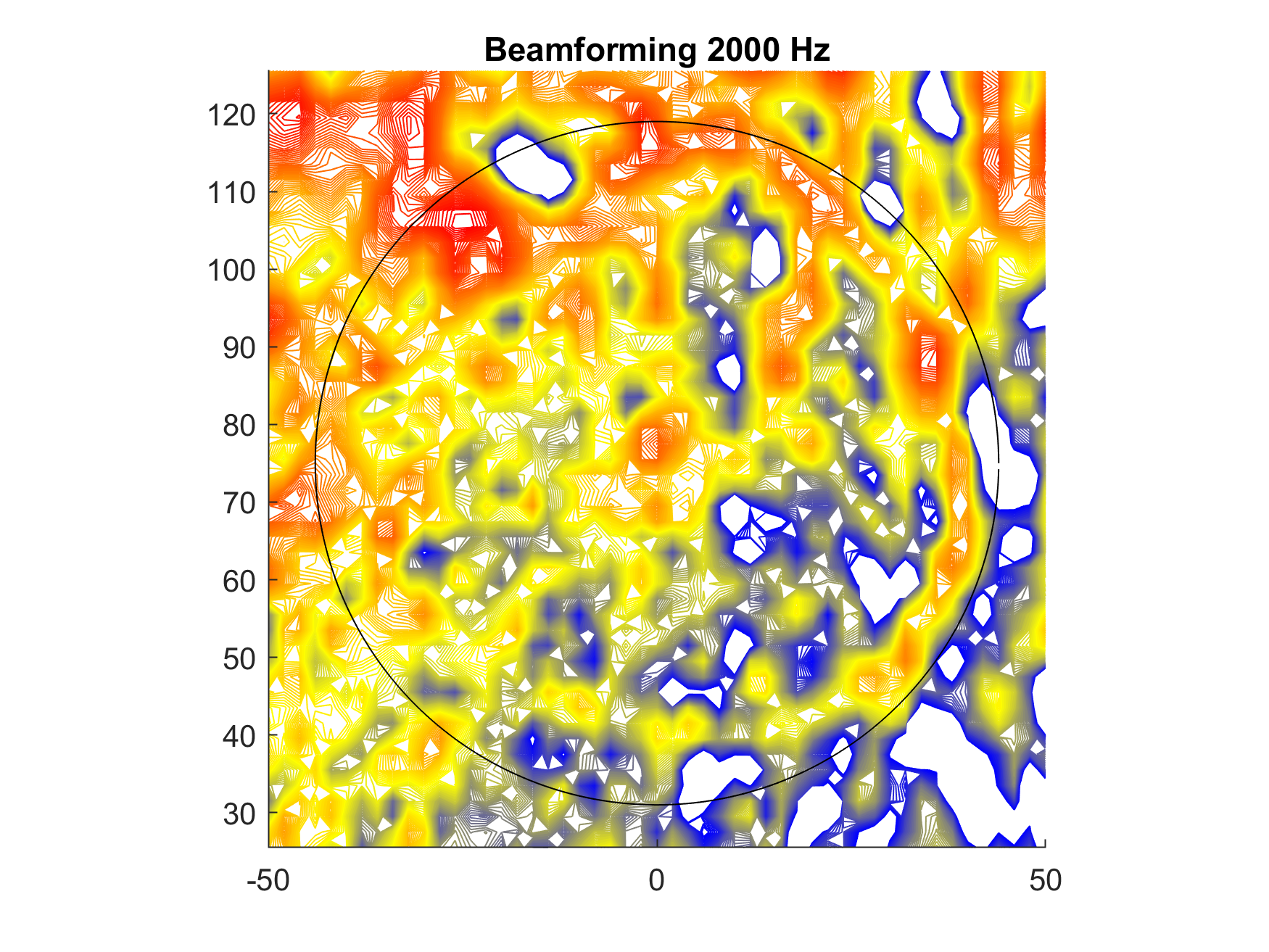}} &   
    {\includegraphics[width=.3\linewidth,trim=66 6 66 6,clip]{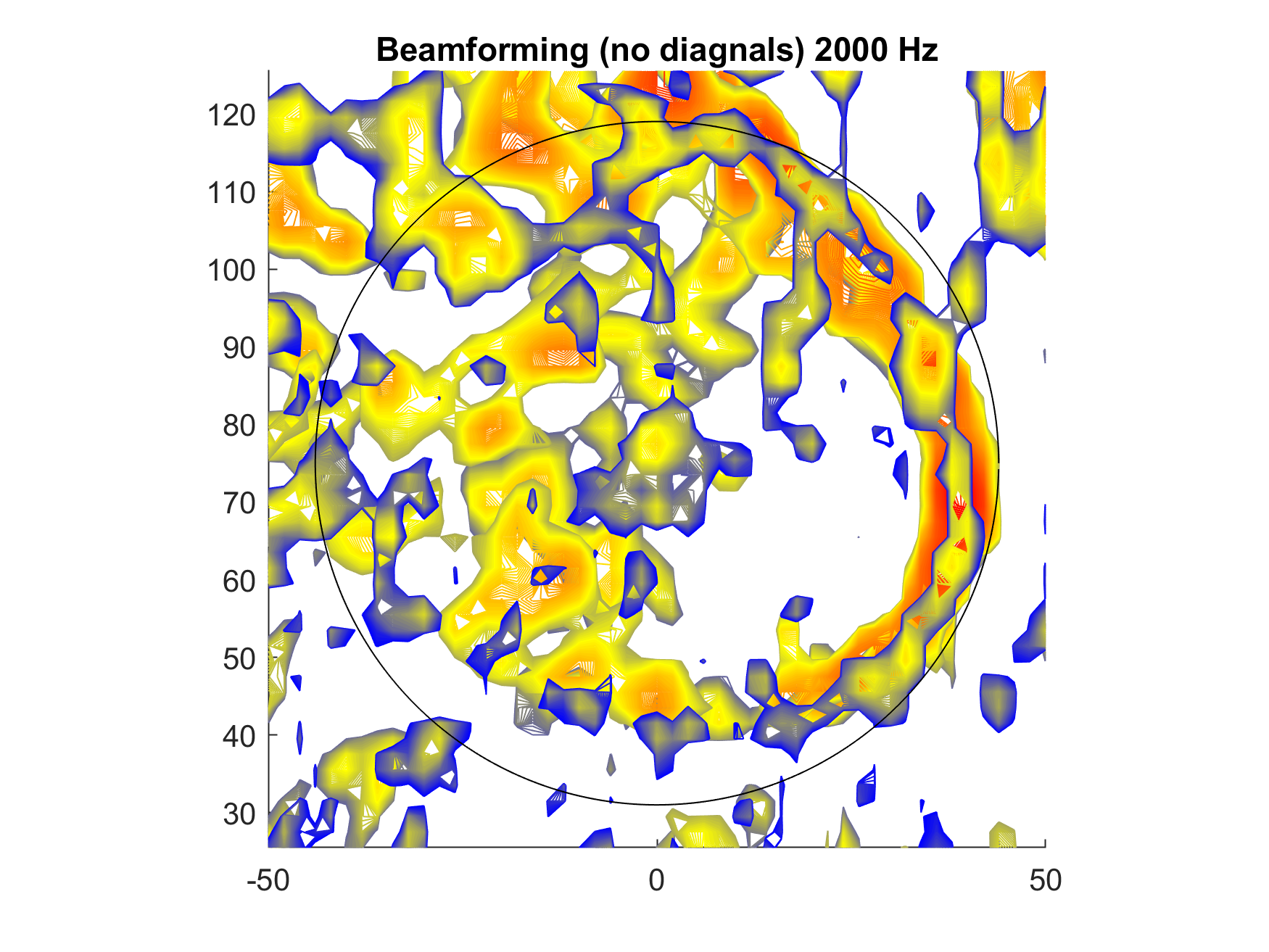}} &   
    {\includegraphics[width=.3\linewidth,trim=66 6 66 6,clip]{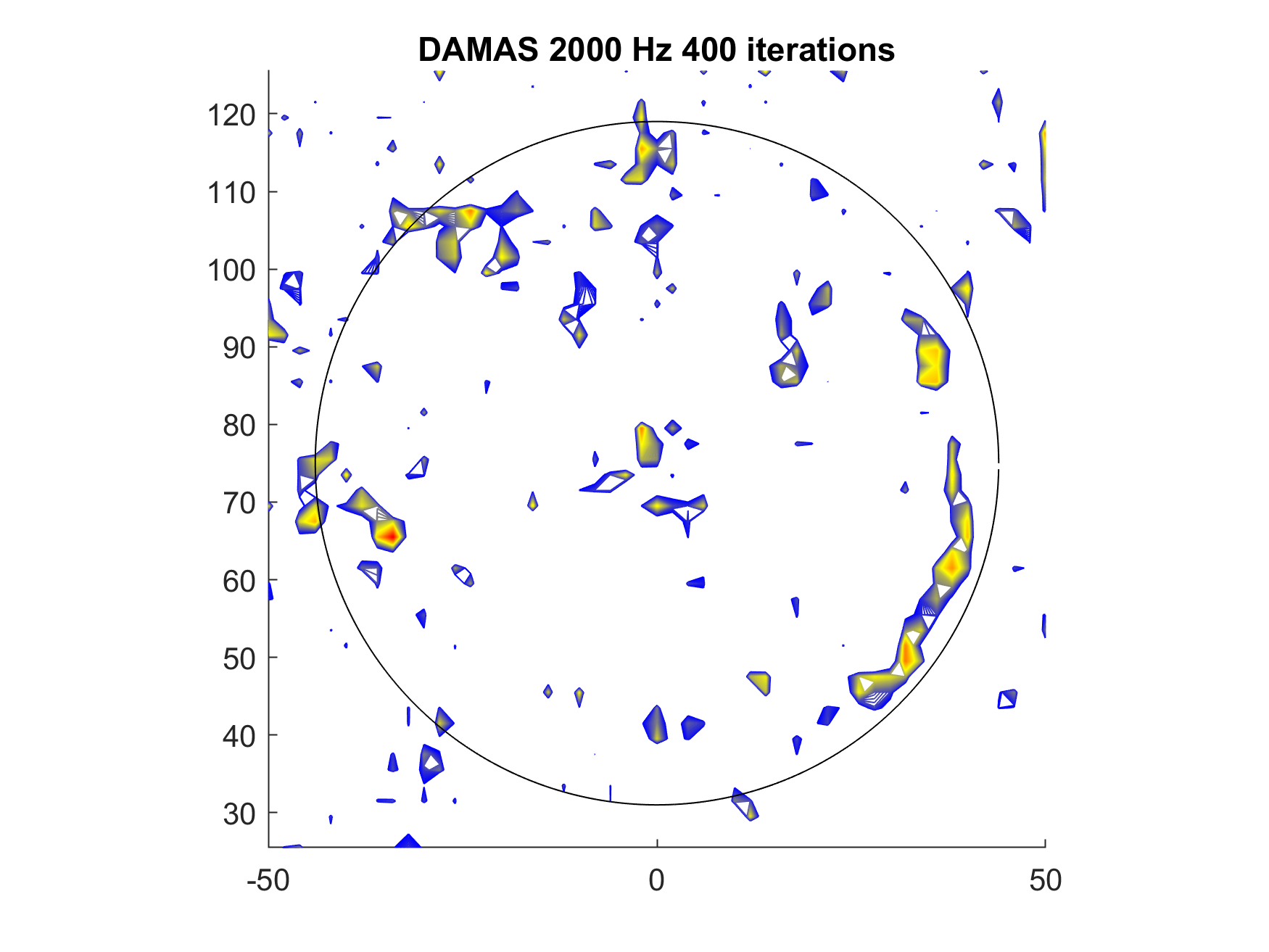}} \\    
\end{longtable}
\caption{Experimental results of sound source localization for a 1.5 MW wind turbine, focusing on 1/3 of the octave band centered from 800 to 2000 Hz. (Left) Traditional beamforming algorithm. (Center) Beamforming with diagonals removal (Right) DAMAS altorithm}
\label{wind_turbine}
\end{figure}

The experimental study focuses on sound source localization of a 1.5 MW permanent magnet direct-drive horizontal-axis wind turbine at a wind farm in Ningxia, China. The turbine's rotor radius is 44.1 m, and the hub height is 75 m. The grid size is set to 100 m in both length and height directions, with a total of 50 grid intervals in each direction, resulting in 2601 grid points in total. Fig.\ref{wind_turbine} displays the results for a frequency range of 800 to 2000 Hz, which covers 1/3 of the octave band. The subfigures in the left and middle columns depict the beamforming algorithm results, with the displayed range being the maximum value and the maximum value -12 dB. The subfigures in the right column show the DAMAS algorithm results, with the displayed range being the maximum value and the maximum value -20 dB.

Based on the experimental study comparing the traditional beamforming algorithm (left) and the diagonal-removed beamforming algorithm (middle) for sound source localization of the 1.5 MW wind turbine, it can be observed that the main lobe size in the noise intensity maps of the traditional beamforming algorithm (left) is significantly larger than that of the diagonal-removed beamforming algorithm (middle), especially at higher frequencies. This indicates that removing the autocorrelation noise using the diagonal-removed approach in the beamforming algorithm helps in understanding the noise distribution of the wind turbine and achieving more accurate sound source localization. Furthermore, it can be observed that in different 1/3-octave frequency bands, the main noise of the wind turbine is generated by the interaction between the blades and the air during blade rotation. As the frequency increases, the sound sources on the blades tend to move radially towards the blade tips, although they are not consistently located at the blade tips. In higher frequency bands, although the DAMAS algorithm (right) yields more noise artifacts in the noise maps, their intensity compared to the main lobe is relatively small. In the noise maps of the traditional beamforming algorithm, when the center frequency of the 1/3-octave band $f$ is greater than or equal to 1000 Hz, false sound sources appear near the highest point of the wind turbine rotor plane, with intensities comparable to those produced by the downward blade motion. This can lead to misunderstandings and misjudgments of the wind turbine's sound source distribution. However, using the DAMAS algorithm to compute the sound source distribution eliminates this interference, which is one of the advantages of DAMAS in terms of accuracy.

\begin{figure}[H]
\begin{longtable}{@{}ccc@{}}
    {\includegraphics[width=.3\linewidth,trim=66 6 66 6,clip]{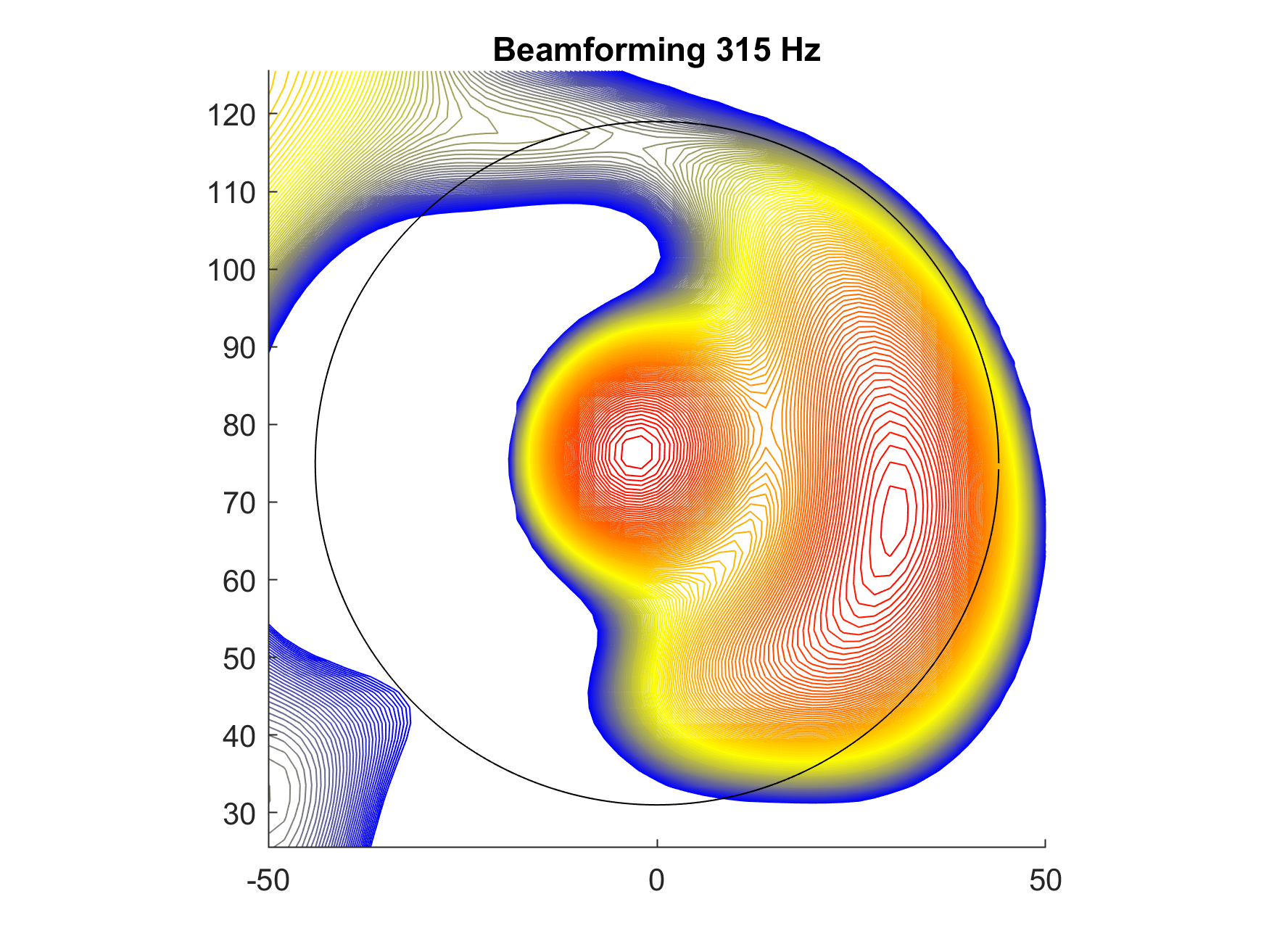}} &
    {\includegraphics[width=.3\linewidth,trim=66 6 66 6,clip]{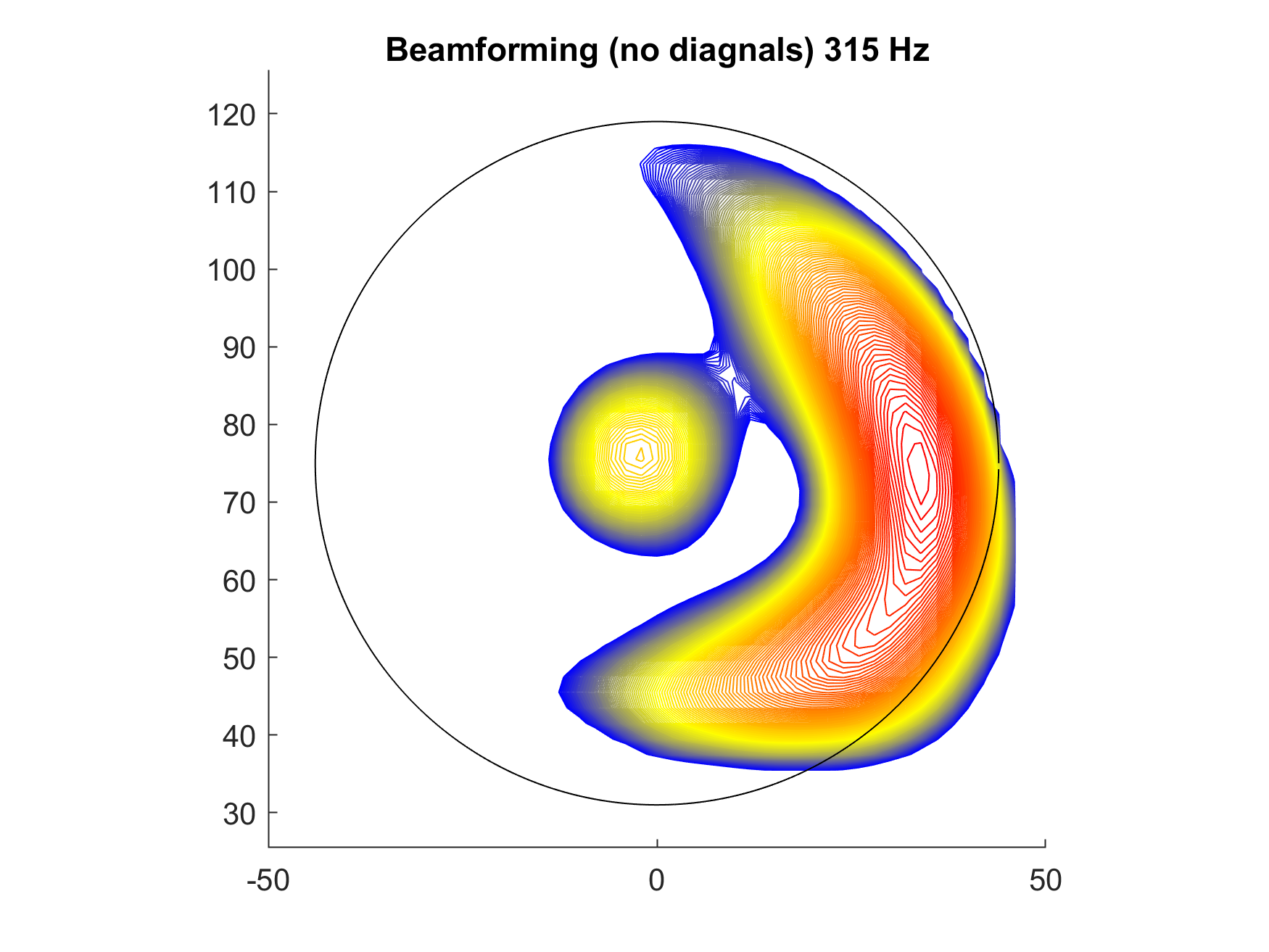}} &
    {\includegraphics[width=.3\linewidth,trim=66 6 66 6,clip]{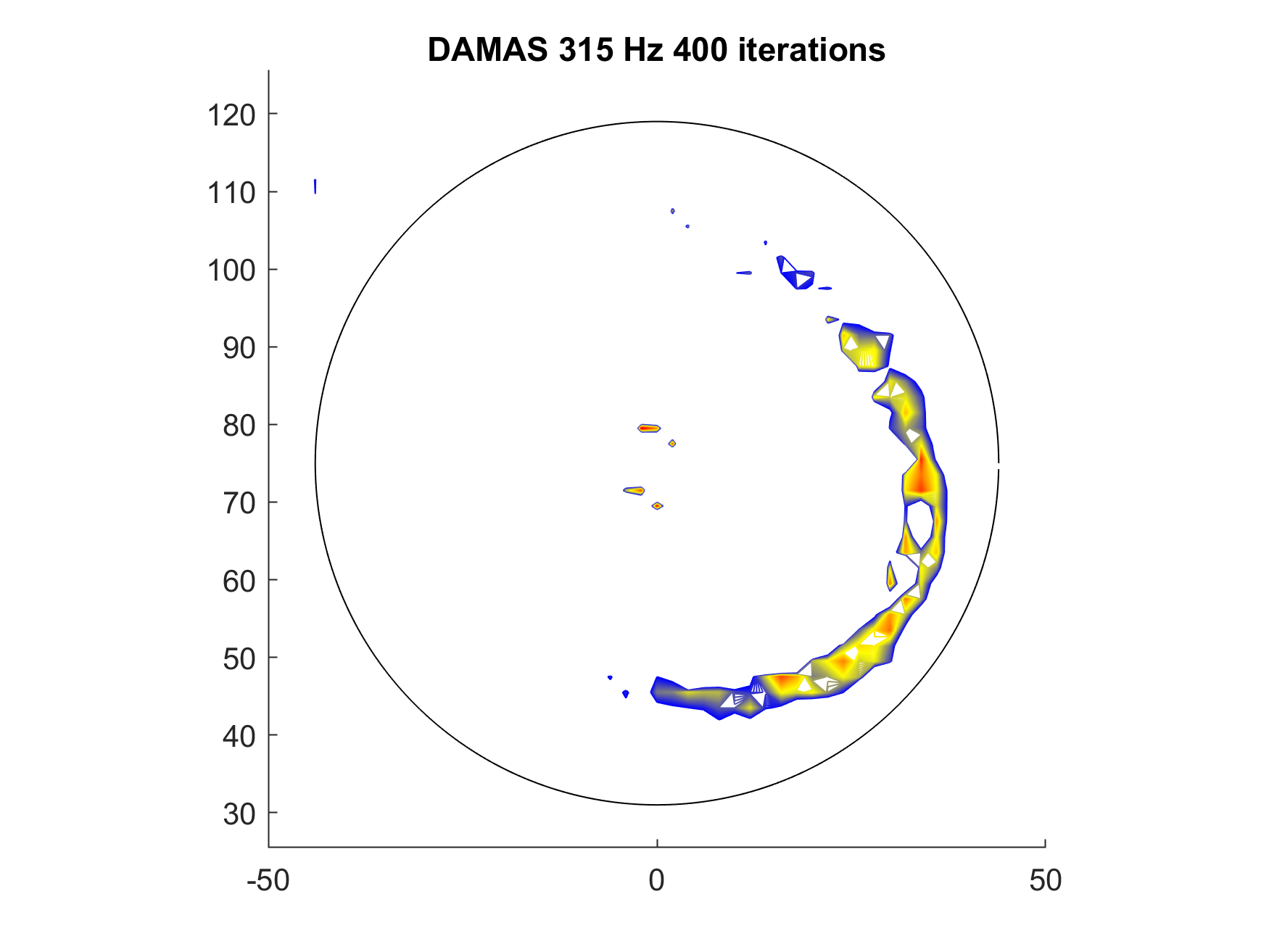}} \\
    {\includegraphics[width=.3\linewidth,trim=66 6 66 6,clip]{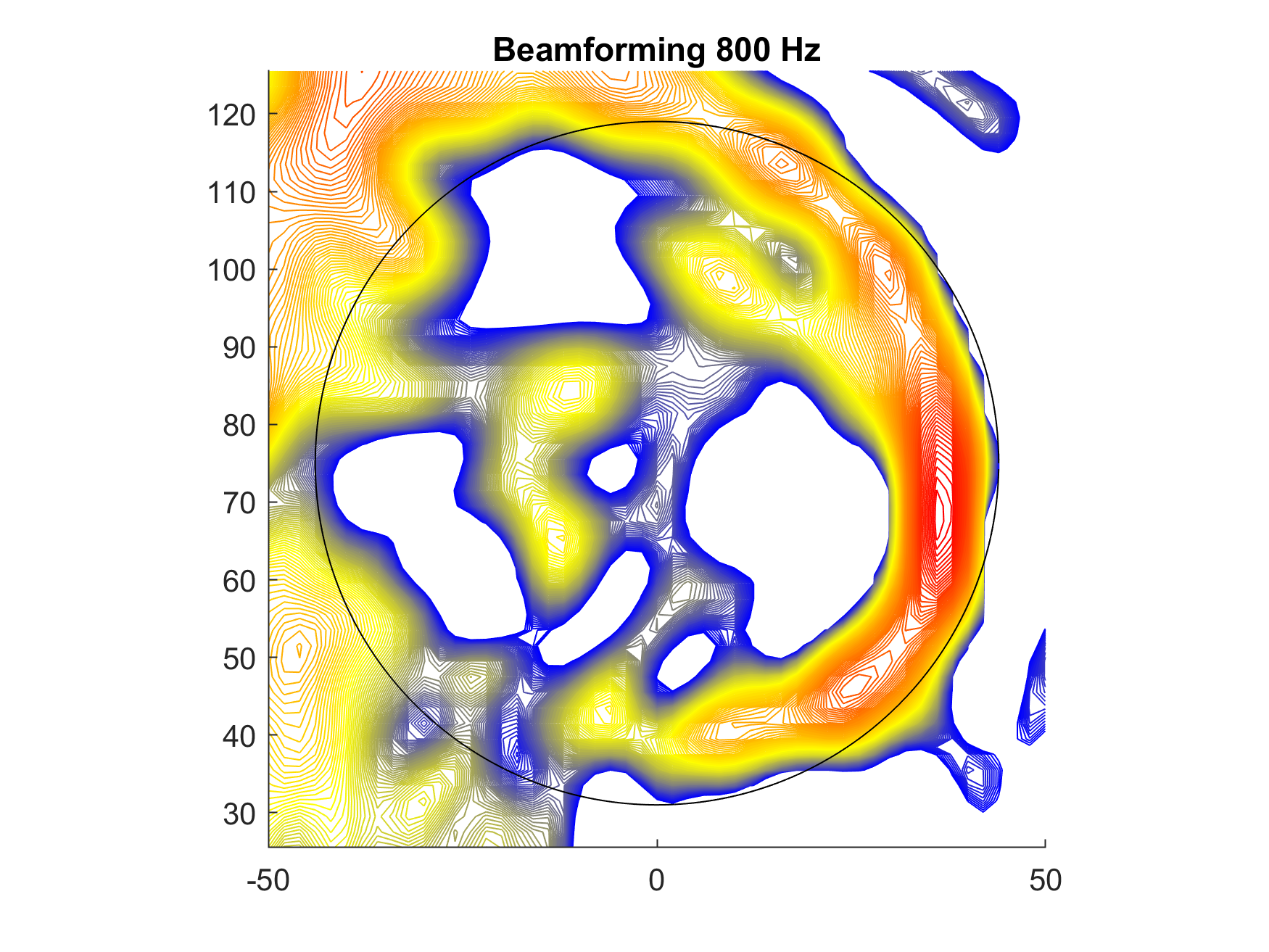}} &    
    {\includegraphics[width=.3\linewidth,trim=66 6 66 6,clip]{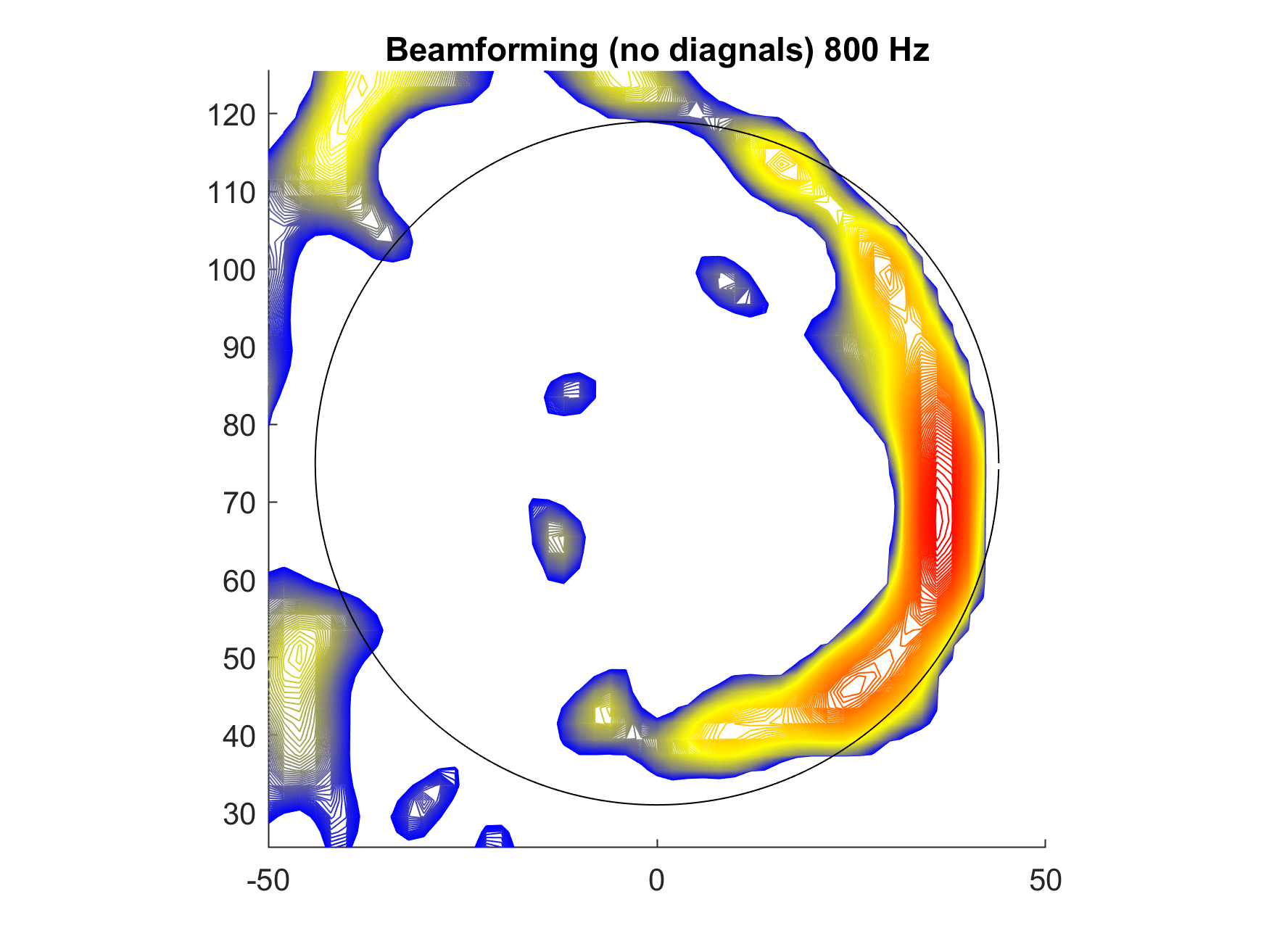}} &
    {\includegraphics[width=.3\linewidth,trim=66 6 66 6,clip]{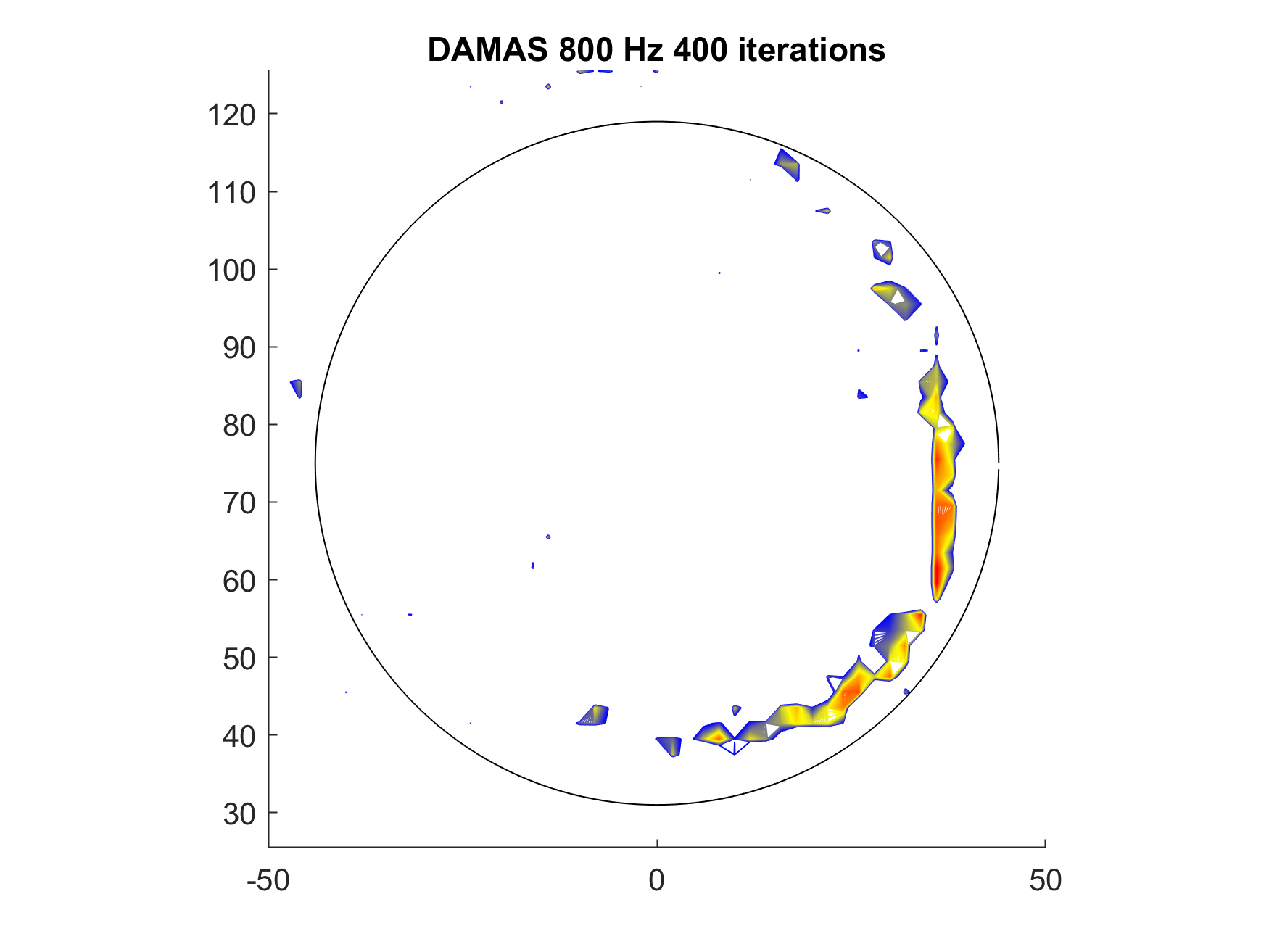}} \\
    {\includegraphics[width=.3\linewidth,trim=66 6 66 6,clip]{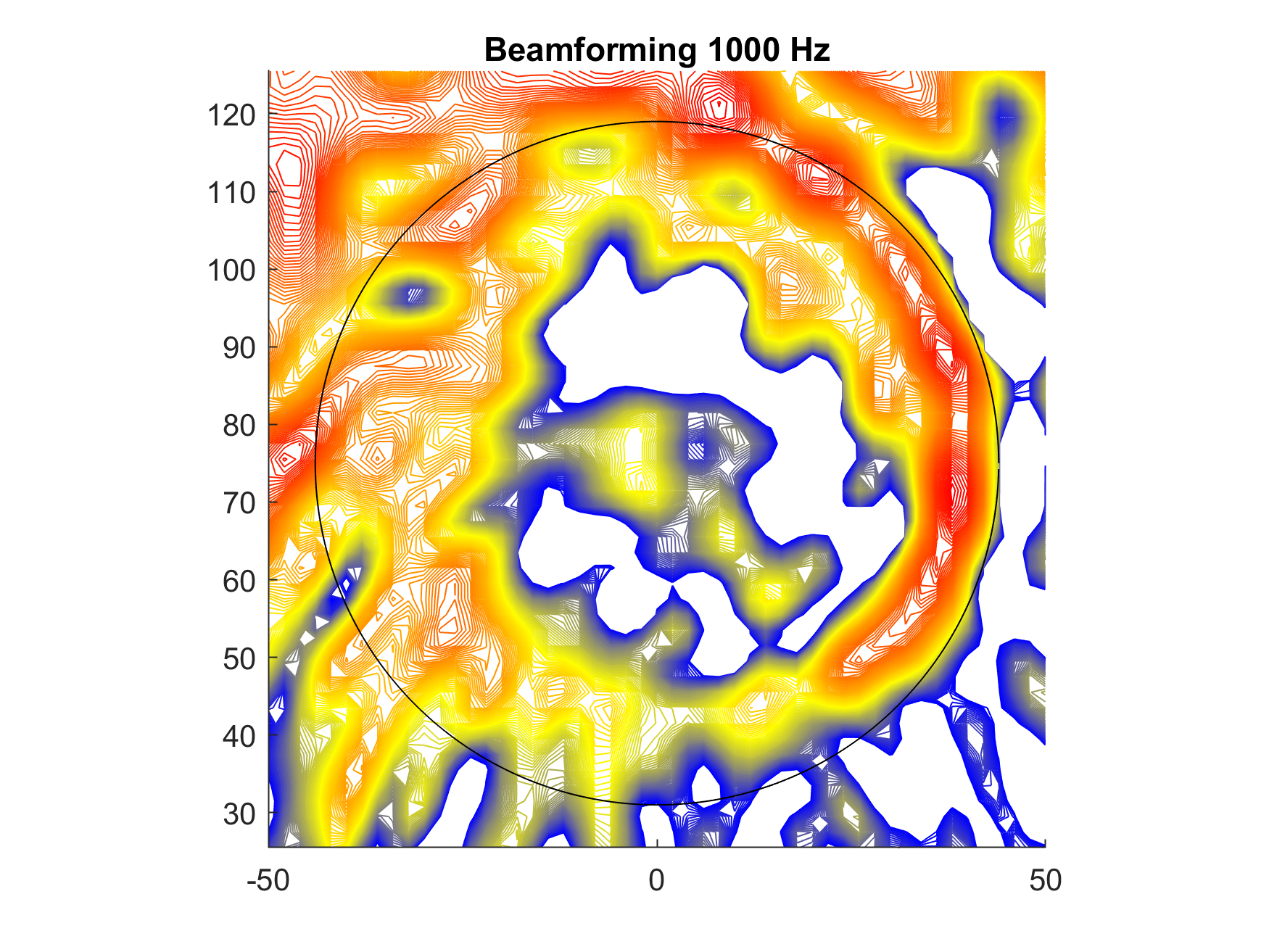}} &   
    {\includegraphics[width=.3\linewidth,trim=66 6 66 6,clip]{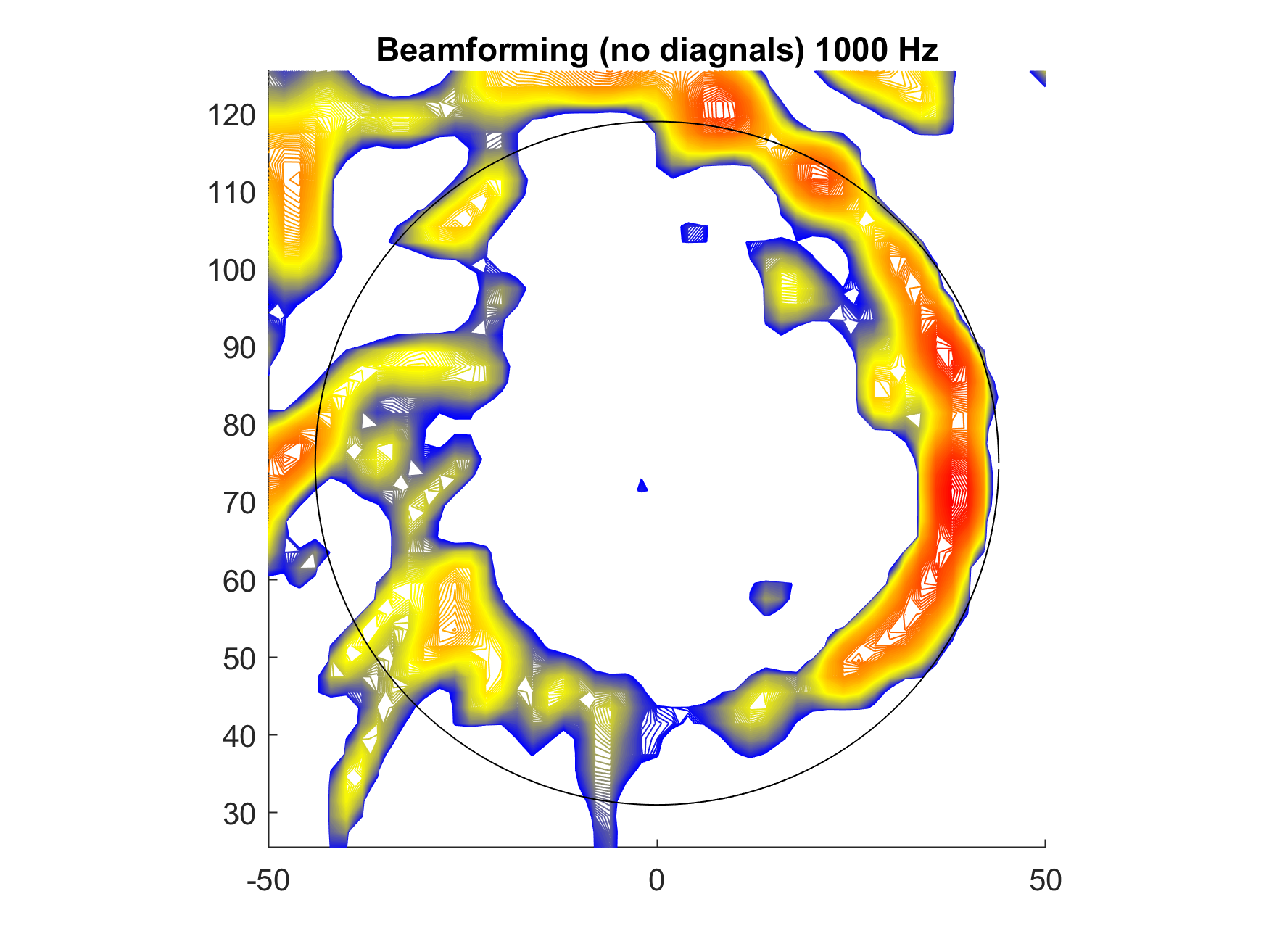}} &
    {\includegraphics[width=.3\linewidth,trim=66 6 66 6,clip]{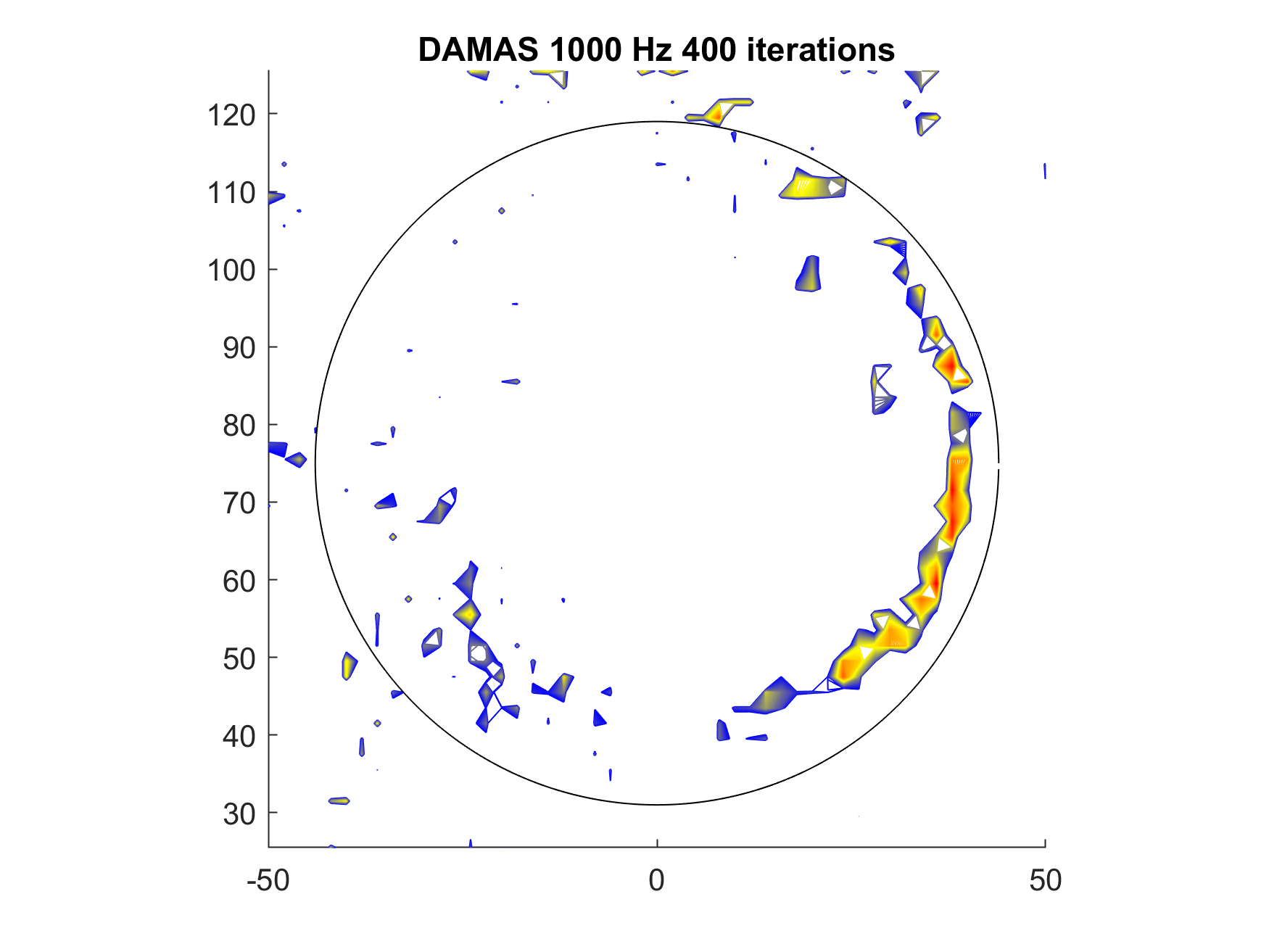}} \\
    {\includegraphics[width=.3\linewidth,trim=66 6 66 6,clip]{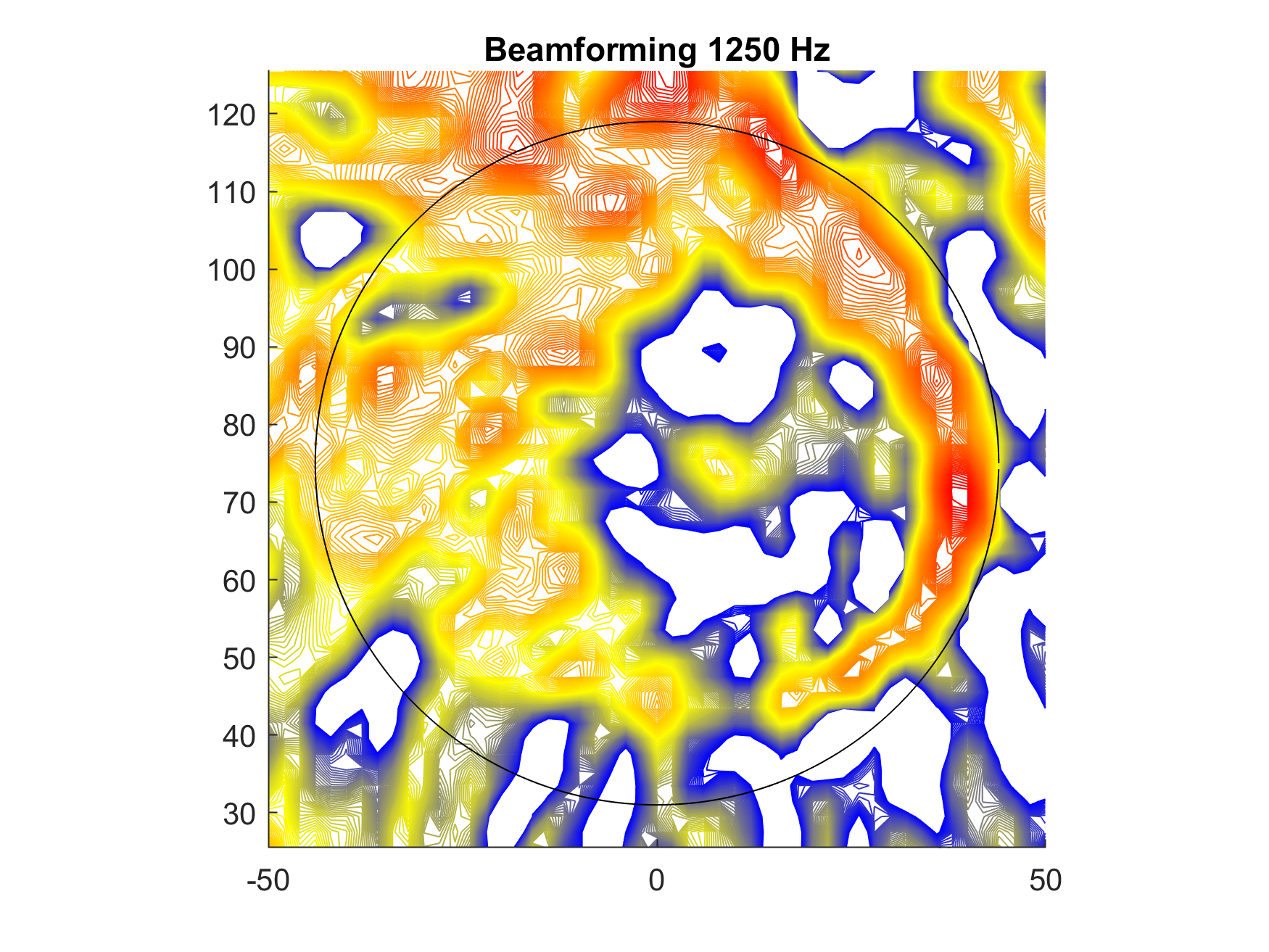}} &   
    {\includegraphics[width=.3\linewidth,trim=66 6 66 6,clip]{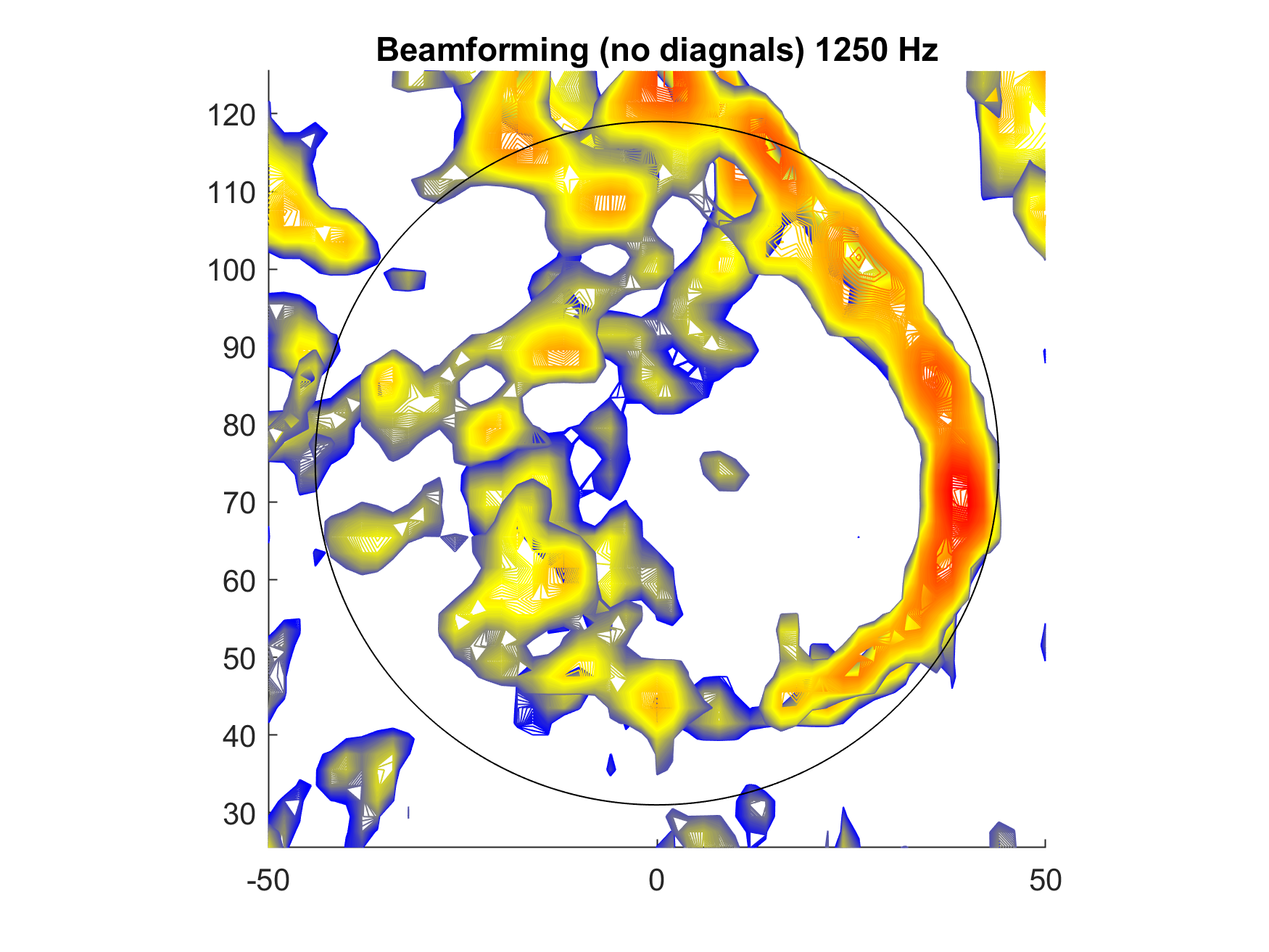}} &   
    {\includegraphics[width=.3\linewidth,trim=66 6 66 6,clip]{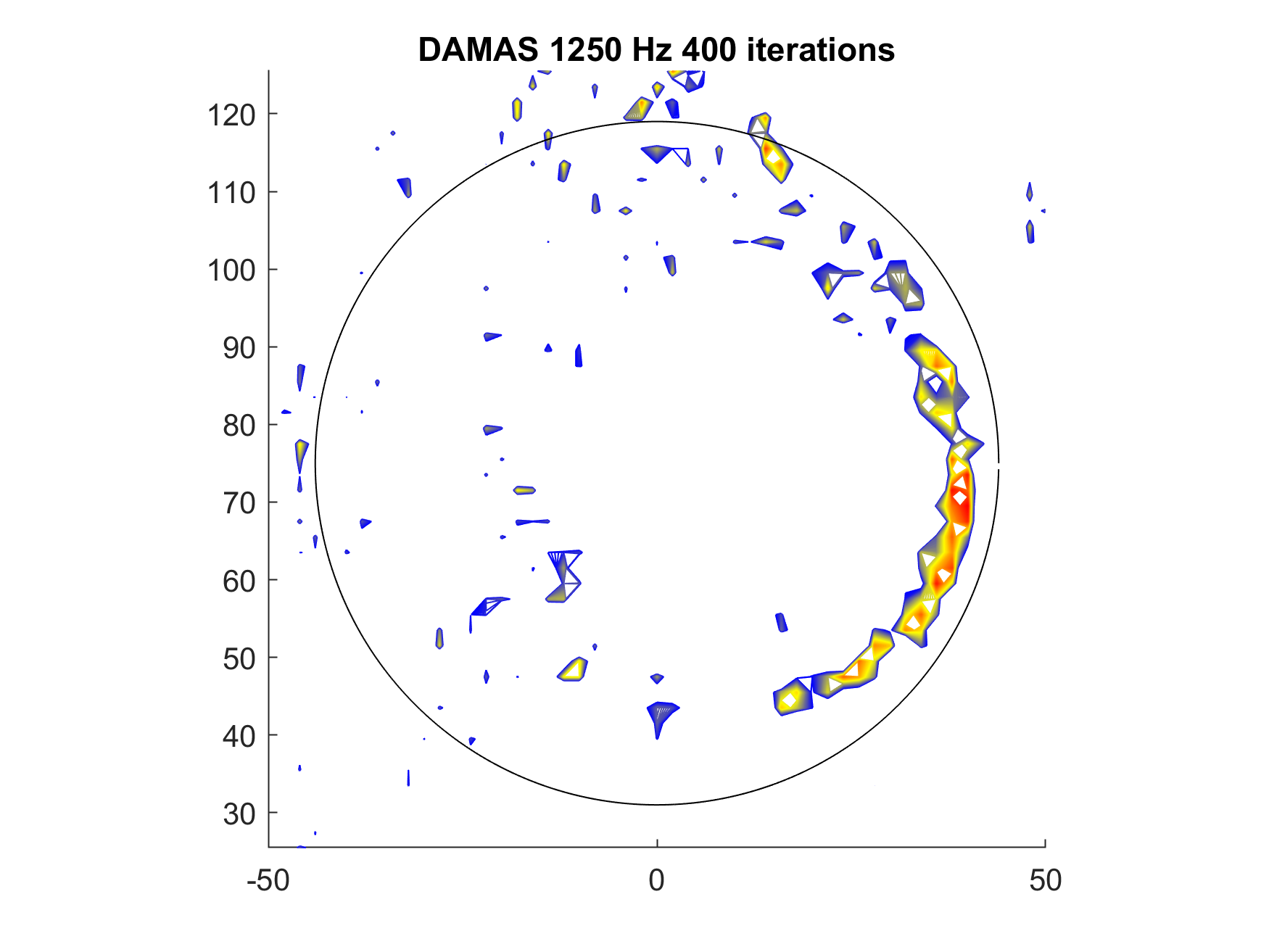}} \\
\end{longtable}
\end{figure}

\section{Conclusions}
\label{conclusions}

The work firstly employed the beamforming algorithm and the DAMAS algorithm to perform simulation-based localization of a 1.5 MW wind turbine for single-frequency point sources, the complex line source resembling the "UCAS" inscription, and the line source downstream of the wing trailing edge. These simulations were performed to verify the effectiveness and accuracy of the DAMAS algorithm. Subsequently, experimental localization studies were conducted on the sound source of the 1.5 MW wind turbine using both the beamforming algorithm and the DAMAS algorithm. By comparing the localization results of the beamforming algorithm and the DAMAS algorithm, the following conclusions can be drawn:

\begin{enumerate}
\item The beamforming algorithm and the power spectrum method can achieve relatively accurate localization of sound sources. However, the interpretation of the results may be ambiguous or even incorrect due to factors such as the number and arrangement of the microphones in the array.

\item The DAMAS algorithm exhibits better accuracy in calculating the position and intensity of the sound source, with smaller main lobe sizes in the resulting intensity maps.

\item The DAMAS algorithm can achieve localization results similar to the initial conditions in simulation-based localization. In experimental localization, more noise artifacts may appear when the frequency is higher. However, compared to the extensive sound source regions observed in the beamforming algorithm, the DAMAS algorithm demonstrates clear advantages.

\item Experimental localization studies can be conducted using a combination of the diagonal-removed beamforming algorithm and the DAMAS algorithm. This approach is beneficial for a more accurate understanding of sound source distribution and for further noise reduction methods.
\end{enumerate}




\bibliographystyle{elsarticle-num-names} 






\bibliography{mybib}

\end{document}